\documentclass[12pt, preprint]{aastex}

\usepackage{amsmath}
\usepackage{natbib}

\newcommand{\XMM}{{\em XMM-Newton }}

\newcommand{\Ch}{{\em Chandra }}

\newcommand{\frd}{Fanaroff-Riley dichotomy }

\def\gappeq{\mathrel{ \rlap{\raise.5ex\hbox{$>$}}
                      {\lower.5ex\hbox{$\sim$}}  } }
\def\lappeq{\mathrel{ \rlap{\raise.5ex\hbox{$<$}}
                      {\lower.5ex\hbox{$\sim$}}  } }

\shorttitle{X_RAY OBSERVATIONS OF RADIO-GALAXY NUCLEI}
\shortauthors{EVANS ET AL.}

\begin{document}

\title{CHANDRA AND XMM-NEWTON OBSERVATIONS OF A SAMPLE OF LOW-REDSHIFT FRI AND FRII RADIO-GALAXY NUCLEI}
\author{D. A. Evans\altaffilmark{1,2}, D. M. Worrall\altaffilmark{1}, M. J. Hardcastle\altaffilmark{3}, R. P. Kraft\altaffilmark{2}, M. Birkinshaw\altaffilmark{1}}
\altaffiltext{1}{University of Bristol, Department of Physics, Tyndall Avenue, Bristol BS8 1TL, UK}
\altaffiltext{2}{Harvard-Smithsonian Center for Astrophysics, 60 Garden Street, Cambridge, MA 02138, USA}
\altaffiltext{3}{School of Physics, Astronomy \& Mathematics, University of Hertfordshire, College Lane, Hatfield, AL10 9AB, UK}

\begin{abstract}

We present spectral results, from {\it Chandra} and {\it XMM-Newton}
observations, of a sample of 22 low-redshift ($z < 0.1$) radio
galaxies, and consider whether the core emission originates from the
base of a relativistic jet, an accretion flow, or contains
contributions from both. We find correlations between the unabsorbed
X-ray, radio, and optical fluxes and luminosities of FRI-type
radio-galaxy cores, implying a common origin in the form of a jet. On
the other hand, we find that the X-ray spectra of FRII-type
radio-galaxy cores is dominated by absorbed emission, with $N_{\rm H}
\gappeq 10^{23}$ atoms cm$^{-2}$, that is likely to originate in an
accretion flow. We discuss several models which may account for the
different nuclear properties of FRI- and FRII-type cores, and also
demonstrate that both heavily obscured, accretion-related, and
unobscured, jet-related components may be present in all radio-galaxy
nuclei. Any absorbed, accretion-related, components in FRI-type
galaxies have low radiative efficiencies.

\end{abstract}


\section{INTRODUCTION}

The physical origin of X-ray emission in the nuclei of radio galaxies
is a topic of considerable debate. It is unclear whether the emission
primarily originates in an accretion flow, or instead has an origin
associated with a parsec-scale radio jet. The
evidence in favor of each interpretation provides motivation to this
paper, which uses spectroscopy of the X-ray cores to
distinguish the two possibilities. On the one hand, it has been suggested that at least a fraction of the nuclear X-ray emission of radio galaxies has an origin at the unresolved base of a parsec-scale radio jet (e.g.,~\citealt{fab84}). The best pieces of evidence in favor of this hypothesis are the observed correlations between the {\it ROSAT} X-ray and VLA radio core fluxes and luminosities measured in the B2 (\citealt{can99}) and 3CRR (\citealt{har99}) samples, supporting a nuclear jet-related origin for at least the soft X-ray emission. \cite{har99} demonstrated that, since the jet-generated radio emission is believed to be strongly affected by relativistic beaming, the flux--flux and luminosity--luminosity correlations would have considerable intrinsic scatter if the X-ray emission were instead to originate in an isotropic accretion flow.

Further constraints on the physical processes present in radio-galaxy
nuclei come from {\it HST} observations
(\citealt{chi99,har00}). Correlations are observed between the radio
and optical luminosities of 3CR FRI-type nuclei, again supporting a
common origin at the base of a parsec-scale jet. However, FRII-type nuclei show a range of behavior at optical wavelengths:
those sources with weak or narrow optical line-emission lie on the
(presumably jet-related) radio--optical luminosity--luminosity
correlation established for the FRI-type sources
(\citealt{chi00b}). However, the optical luminosities of broad-line
FRII-type sources lie significantly above this trendline, and
\cite{chi00b} interpret these sources as being oriented to the
observer such that significant optical emission from an accretion disk is
observed, consistent with predictions from AGN-unification schemes
(e.g.,~\citealt{ant93,up95}). Additional evidence in favor of
jet-related X-ray emission in the nuclei of radio galaxies is found when
considering the multiwavelength spectral energy distributions (SED) of
these sources. As demonstrated by \cite{cap02}, \cite{chi03}, and
\cite{pel03}, the nuclear SED of certain FRI-type radio galaxies may
be modeled by synchrotron and synchrotron self-Compton emission from
the base of a relativistic jet.

There is, however, an alternative interpretation of the origin of
nuclear X-ray emission: that it is emission from an accretion
flow. Observations in favor of this model include the detection of
broadened Fe K$\alpha$ lines and short-timescale ($\sim$ks) variability, both of
which were used to argue for a physical origin in the inner regions of an accretion
flow (\citealt{gli03,gli04}). In a recent study of a sample of
FRI-type radio-galaxy nuclei, \cite{don04} showed that several sources
with strong optical jet emission are not accompanied by strong X-ray
emission, which may suggest a different physical origin for the
two. Further, \cite{mer03} argued that the observed correlations
between the luminosities of radio and X-ray emission in radio-galaxy
cores may be part of a `fundamental plane', linking the radio and
X-ray emission with black hole mass in both radio-loud and radio-quiet
AGN. They suggest that the X-ray emission in all these sources has an
origin in the form of an accretion flow, although they cannot rule out
a significant contribution from the jet to the X-ray emission of the radio-loud sources.

In addition to the debate surrounding the origin of the X-ray
emission, there has been some controversy over whether physical
differences exist in the accretion-flow modes of FRI- and FRII-type
sources, and over the presence and r\^{o}le of obscuring tori in these
sources. It is unclear whether the \frd observed on large (kpc--Mpc)
scales (\citealt{fr74}) is due to the primarily {\it extrinsic} impact
of the hot-gas environment on the jet propagating through it
(e.g.,~\citealt{bic94,bic95,gkw00}), or rather ensues from {\it
intrinsic} differences in the structure of the accretion flow and/or
torus (see discussion by \citealt{gkw00}). There is growing evidence
(e.g.,~\citealt{rey96,don04}) to suggest that FRI-type radio-galaxy
nuclei possess radiatively inefficient, possibly advection-dominated
(e.g.,~\citealt{nar95}) accretion flows, rather than standard,
geometrically thin disks. It has also been claimed from the low
intrinsic absorption measured at X-ray wavelengths (\citealt{don04}),
together with the high optical-core detection rate (\citealt{chi99}),
that an obscuring torus, required by AGN-unification schemes, is {\it
absent} in FRI-type radio galaxies. However, as noted by
\cite{har02b},~\cite{wor03b}, and \cite{cao04}, if the optical and
X-ray emission is dominated by a jet and occurs on scales larger that
of a torus, then one cannot comment directly on the presence or
absence of the torus using the X-ray absorption and optical reddening properties alone.

As the orientation-dependent effects of relativistic beaming and the
putative obscuring torus are expected to play a large part in
determining the observed properties of a radio-galaxy nucleus, it is
important to select sources based on their low-frequency (and hence
isotropic) emission characteristics, such as in the 3C and 3CRR
samples. In this paper, we present the results of a \Ch and \XMM
spectral analysis of a sample of the nuclei of 22 radio galaxies at $z
< 0.1$, 19 of which are from the 3CRR catalogue, with the remaining
sources, 3C~403, 3C~405 (Cygnus~A), and Centaurus~A, included due to
their high quality spectra. The high spatial resolution of \Ch means that it is
possible to disentangle confusing kpc-scale jet emission from that of
the core, while the large collecting area of \XMM allows tight
constraints to be placed on spectral parameters. Each observatory
covers a sufficiently large energy range that both soft unobscured
X-ray emission and hard obscured emission, possibly viewed through a torus, may be observed, providing direct tests of AGN-unification models.

This paper is organized as follows. Section 2 contains a description
of the data and a summary of their analysis. The results from fitting
the spectra of the nuclei of each source are presented in Section
3. In Section 4, we discuss the X-ray, radio, and optical flux and
luminosity correlations. In Section 5, we discuss the presence that an
obscuring torus is present in FRI-type radio galaxies, while in
Section 6 we interpret the observed differences in the X-ray nuclei of FRI-
and FRII-type sources. We end with our conclusions in Section 7. All results presented in this paper use a cosmology in which $\Omega_{\rm m, 0}$ = 0.3, $\Omega_{\rm \Lambda, 0}$ = 0.7, and H$_0$ = 70 km s$^{-1}$ Mpc$^{-1}$. When distinguishing between different model fits to the data, we present $F$-statistic results, although we note that this method may be unreliable in such circumstances (\citealt{pro02}). We adopt thresholds of 95 and 99.9 per cent for marginally and highly significant improvements in the fit, respectively. Errors quoted in this paper are 90 per cent confidence for one parameter of interest (i.e., $\chi^2_{\rm min}$ + 2.7), unless otherwise stated.

\section{OBSERVATIONS AND ANALYSIS}

Out of a total of 35 radio galaxies at $z < 0.1$ in the 3CRR
catalogue (excluding the starburst galaxy 3C 231), 19 have been observed with \Ch or {\it XMM-Newton}. In total, 16 \Ch and 5 \XMM
observations of the sources are available at the time of writing, with
the majority taken from the public data archives, together with some
proprietary GO and GTO data. We compare results with those for Centaurus~A (\citealt{evans04}), 3C~403
(\citealt{kra05}), and 3C~405 (Cygnus~A). Centaurus~A is the nearest
and best-studied AGN, while 3C~403 and 3C~405 are nearby FRII-type
sources, and so each provides a useful comparison with the other
sources in our sample. The main properties of the sources, including their redshift, Fanaroff-Riley classification,
optical source type [Low-Excitation Radio Galaxy (LERG), Narrow-Line
Radio Galaxy (NLRG), or Broad-Line Radio Galaxy (BLRG)], and V-band
apparent magnitude, are given in Table~\ref{sources_overview}. The redshifts quoted
are taken from the online 3CRR
catalogue\footnote{http://www.3crr.dyndns.org/}, compiled by
M. J. Hardcastle based on data from \cite{lai83} and with updates
collated by Laing, Riley, and Hardcastle. The classification of the
excitation properties of the sources follows
\cite{lai94} and~\cite{jr97}, who define high-excitation objects as
having [OIII]/H$\alpha > 0.2$ and equivalent widths of [OIII] $>
3$\AA.

The \Ch data were reprocessed using {\sc CIAO} v3.1 with the
{\sc{CALDB}} v2.2.8 calibration database to create a new level 2
events file with grades 0, 2, 3, 4, 6, afterglow events preserved, and
the 0.5-arcsec pixel randomization removed. To check for intervals of
high particle background, light curves were extracted for the chip
upon which the source was located, excluding the source itself and any
other noticeable point sources. The light curves were filtered based
on the 3$\sigma$-clipping method of the user-contributed {\sc
analyze\_ltcrv} script, available from the {\sc CXC}
website\footnote{http://cxc.harvard.edu}.

The \XMM data were reprocessed using {\sc SAS} version 5.4.1, and
calibrated event files were generated using the {\sc EMCHAIN} and {\sc
EPCHAIN} scripts, with the additional filtering criteria of selecting
events with only the PATTERN$\leq$4 and FLAG=0 attributes. Periods of
high particle background were screened by extracting light curves from
the whole field of view, excluding a circle centered on the source,
and selecting only events with PATTERN=0 and FLAG=0 attributes and for
an energy range of 10--12 keV (MOS) cameras and 12--14 keV (pn).

For the analysis of nuclear emission, it is important to minimize the
contribution of contaminating extended emission to the nuclear
spectrum, meaning that \Ch observations of the sources are
desirable. We extract on-source spectra from a small-radius circle
(typically 2.5 pixels or 1.23$''$ in the case of {\it Chandra}, and
35$''$ in the case of {\it XMM-Newton}) and
use local background subtraction. We take care to exclude unrelated
contaminating sources and resolved kiloparsec-scale jet emission. The
fluxes and luminosities quoted in this paper are corrected for the
(generally small - typically $\sim$10--15\%) fraction of missing counts
that are the result of using a small aperture and the local background
subtraction.

Spectra were extracted and calibration files generated using the
standard {\sc CIAO} and {\sc SAS} scripts, {\sc psextract} and {\sc
evselect}, respectively. Spectral fitting was performed on data
grouped to a minimum of 25 counts per bin over an energy range 0.5--7
keV ({\it{Chandra}}) and 0.5--10 keV ({\it{XMM-Newton}}). Models were
fitted to the background-subtracted source spectra, increasing
in complexity until an adequate fit was achieved. In cases where two
components of similar power-law spectral indices but different
absorptions give the best fit to a spectrum, we cannot rule out the
alternative of a single power-law index and a range of absorptions. The results of the
spectral fitting were checked for their consistency, either by
comparing them with previously published work, or, where possible, by
intercomparing results from {\it Chandra} and {\it XMM-Newton}. In
cases where the addition of a small-scale thermal component to the
nuclear spectra of \Ch data was necessary, a consistency check was performed
between the numbers of thermal counts found spectrally and extended
counts found spatially via fitting a point source and $\beta$-model
convolved with the PSF to the radial surface-brightness profile.

Pileup can be a concern when analyzing the X-ray spectrum of even a
moderately bright point source, especially with \Ch data. A
consideration of the count rates and an inspection of the spatial
distribution of \Ch images filtered solely for `afterglow' events (see
discussion by \citealt{evans05}) showed that observations of three
sources, Centaurus~A, NGC~6251 and 3C~390.3, are significantly affected by pileup
(pileup fraction $>10$\%). The piled spectrum of Centaurus~A is
extensively discussed by \cite{evans04}. The
nuclear spectrum of NGC~6251 may be recovered by using an annular
extraction region and therefore sampling only the wings of the
PSF (\citealt{evans05}). For 3C~390.3, the source flux is sufficiently high that the
nuclear spectrum of the source may be extracted from the frame
transfer streak, using the method of
\cite{mar05} to correct for the fraction of events that occurred in
the frame transfer streak.

\section{RESULTS}

In Table~\ref{3crr_spectral results}, we give the results of our
analyses of each source. It is clear that no single model provides an
adequate fit to every spectrum, with some sources having essentially
no intrinsic absorption, and others having absorbing columns in excess
of $10^{23}$ atoms cm$^{-2}$, together with fluorescent Fe K$\alpha$
line-emission. We note that thermal emission from an extended component is
sometimes not fully subtracted by local background subtraction. A range of power-law indices is also observed:
best-fitting values range from 1.47 to 2.37. Further, the 2--10 keV
intrinsic (unabsorbed) luminosities of the primary power law
components span five orders of magnitude, with values ranging from
$2\times10^{39}$ to $3\times10^{44}$ ergs s$^{-1}$. Again, we note
that for spectra for which the best fitting model is two components of
similar power-law spectral indices but different absorptions, we
cannot rule out the alternative of a single power-law index and a range of absorptions.

In Table~\ref{3crr_xro_flux_lum_tab}, we show the intrinsic absorption
and unabsorbed 1-keV flux and luminosity densities of each X-ray
spectral power-law component in the best-fit model for the sources. In addition, we
give the flux and luminosity densities of the unresolved cores of each
source, measured at 5 GHz with the VLA and with {\it HST} at red wavelengths (typically using the
F702W filter). The 5-GHz VLA flux and luminosity
densities are taken from the online 3CRR
catalogue\footnote{http://www.3crr.dyndns.org/} and references
therein. The majority of the {\it HST} optical values are taken from
\cite{har00}, who extracted the unresolved core flux densities from
circular extraction regions centered on the source. The {\it HST}
values are dereddened for Galactic absorption only. These values
were checked for their consistency in an independent analysis
(O. Shorttle, private communication), and in several cases this
analysis provided values for sources not studied by
\cite{har00}. An independent analysis of the optical nuclei of
radio galaxies sources was performed by \cite{chi99}; the values
we quote in this paper generally agree with those of \cite{chi99} to
within a factor of 2, which is reasonable given the systematic
uncertainties in background subtraction. Not all sources whose X-ray spectra we have analyzed
have accompanying {\it HST} observations, either due to them not being
observed, or due to other complications, such as the presence of
foreground stars, or the mispointing of the telescope. Note that the
X-ray, radio, and optical observations are not contemporaneous.

\section{ORIGIN OF X-RAY EMISSION}

\subsection{Distribution of intrinsic absorption}

Figure~\ref{ch7_nhhisto} shows a histogram of the intrinsic absorption
associated with the dominant component (in terms of unabsorbed 1-keV
flux density) of X-ray emission in each of the sources. Note that some {\it sources} have two {\it components} of X-ray emission. 3C~388 is not included on this plot, as the detection of its nucleus is somewhat uncertain, and its intrinsic absorption is unconstrained (see Appendix~A). The distribution of core intrinsic absorption is essentially bimodal, with 9 sources having no intrinsic absorption detected and 7 having absorption in excess of $10^{23}$ atoms cm$^{-2}$. One might expect absorbing columns of $N_{\rm H} \gappeq 10^{23}$ atoms cm$^{-2}$ when X-ray emission is surrounded by a dense, dusty structure, such as the putative torus (e.g., \citealt{up95}).

From Figure~\ref{ch7_nhhisto}, it is clear that X-ray emission
components in FRI-type radio galaxies tend to have much lower
intrinsic absorption than FRII-type radio galaxies. This may suggest
an intrinsic difference in the nuclear emission characteristics of
FRI- and FRII-type sources, a subject that we shall return to in
Section~\ref{ch7_dichotomy}. Note that the FRII-type radio galaxy 3C~390.3 is cross-hatched in Figure~\ref{ch7_nhhisto}. This is because it is a broad-line source, which means that, in the unified scheme of AGN, it is likely to be oriented to the observer such that the inner regions of its AGN are exposed, unlike the other FRII-type sources.

\subsection{The radio core -- X-ray core correlation}
\label{ch7_rx}

Figure~\ref{ch7_rxlumflux}a shows a plot of the unabsorbed 1-keV X-ray luminosity
density against the 5-GHz VLA unresolved radio core luminosity density
of each X-ray component of emission of all the sources presented in
Table~\ref{3crr_xro_flux_lum_tab}, while Figure~\ref{ch7_rxlumflux}b shows
a plot of the unabsorbed 1-keV and 5-GHz flux densities. The
components are separated into those with intrinsic absorption less
than $5\times10^{22}$ atoms cm$^{-2}$ ({\it circles}) and those with
intrinsic absorption greater than this value ({\it triangles}). From
Figures~\ref{ch7_rxlumflux}a and~\ref{ch7_rxlumflux}b, it is evident that the
X-ray emission separates into two distinct ``bands'': in other words,
the unabsorbed X-ray luminosities and fluxes of components with high
intrinsic absorption (greater than $5\times10^{22}$ atoms cm$^{-2}$)
tends to lie significantly above than those components with low
intrinsic absorption (less than $5\times10^{22}$ atoms cm$^{-2}$). For
components with $N_{\rm H} < 5\times10^{22}$ atoms cm$^{-2}$, we see
no correlation between X-ray luminosity and intrinsic absorption. Most interestingly, the X-ray emission of all FRII-type sources is dominated by components of high intrinsic absorption, a subject we shall return to in Section~\ref{ch7_dichotomy}.

Figure~\ref{ch7_nhhisto} shows that the majority of X-ray components
below the arbitrary $5\times10^{22}$ atoms cm$^{-2}$ cutoff we have
adopted have essentially no intrinsic absorption. The highest
intrinsic absorption below the cutoff is $\sim4\times10^{22}$ atoms
cm$^{-2}$, which is associated with the second power law in
Centaurus~A, a galaxy noted for its strong and complex dust lane. The
next two highest intrinsic absorptions measured below the cutoff are
in 3C 83.1B and 3C 296, two galaxies observed to possess highly
inclined dusty disks (\citealt{chi99}). In Section~\ref{ch7_optical}
we present evidence that the intrinsic absorption associated with these components is due to gas in the host galaxy. The intrinsic absorption of X-ray components above the cutoff ranges from $\sim$1--6$\times10^{23}$ atoms cm$^{-2}$ and is unlikely to be due to gas in the host galaxy, and is more plausibly associated with denser gas and dust close to the nuclei, such as a dusty torus.

In Figures~\ref{ch7_rxlumflux}a and~\ref{ch7_rxlumflux}b, we identify the
broad-line FRII-type radio galaxy 3C~390.3 which is likely to be oriented to the observer such that the inner regions of the AGN are exposed. We also identify 3C~388, the only low-excitation FRII-type radio galaxy in the sample. As discussed in Appendix~A, its spectrum is uncertain, and may be modeled with either no intrinsic absorption, or an intrinsic absorption of $\sim$$10^{23}$ atoms cm$^{-2}$, consistent with that of the other FRII-type radio galaxies we analyzed. We therefore include {\it two} data points for 3C~388, to illustrate this. Finally, we note that the X-ray spectrum of 3C~321 is somewhat uncertain, meaning that no tight constraints can be placed on the strength of its heavily absorbed emission, while the poor signal-to-noise spectrum of 3C~449 means that it is only possible to place an upper limit on its continuum emission.

In summary, the observed separation of those components with low and high intrinsic absorption may suggest that different physical emission processes are responsible for each. In what follows, we investigate this hypothesis further.

\subsection{Components with low intrinsic absorption: A jet origin}

In this section, we consider those components of X-ray emission with
low intrinsic absorption ($N_{\rm H} < 5\times10^{22}$ atoms
cm$^{-2}$), excluding the broad-line radio galaxy 3C~390.3. The mean photon index for these
components is $1.88\pm0.02$. A strong
correlation is apparent, in both the X-ray/radio luminosity-luminosity
and flux-flux plots (Figure~\ref{ch7_rxlumflux}). Analysis with the astronomical survival analysis
package ({\sc ASURV} --- \citealt{lav92}) implementation of the
modified Kendall $\tau$ algorithm, taking into account the censoring
of the X-ray data, shows that the luminosity-luminosity and flux-flux
correlations are significant at the 99.8\% and 99.95\% significance
levels, respectively. That the correlation in the flux-flux
relationship is significant gives confidence that we are not simply
seeing an artificial redshift-induced artifact in the
luminosity-luminosity relationship. Some scatter still exists in the
distribution, although this may simply be due to variability and the
non-contemporaneous nature of the X-ray and radio data (see, e.g.,
\citealt{evans04} and \citealt{evans05} for discussions related to
Cen~A and NGC~6251).

In order to provide a quantitative analysis, we determined the slope of the core flux-flux and luminosity-luminosity plots using linear regression. Performing linear regression of this dataset is hampered by the fact that some of the data points are upper limits, rather than detections, and so we use the Buckley-James method of linear regression, as implemented in {\sc ASURV}, which takes into account censored data. We calculated the best-fit slope by taking the bisector of the two lines of best fit obtained by the Buckley-James regression of each variable (i.e., the radio and X-ray flux/luminosity) on each other. Taking the bisector is important so that one quantity is not privileged over the other by being treated as the independent variable in the analysis (see discussion by ~\citealt{har99}). For the luminosity-luminosity plot, we find the best-fitting slope to be 0.91 $\pm$ 0.17, and for the corresponding flux-flux plot, we find the best-fitting slope to be 1.06 $\pm$ 0.16.

The observed correlation implies a physical relationship between the
X-ray emission and jet-generated radio-core emission of sources with
components of low intrinsic absorption. This confirms previous work
(e.g.,~\citealt{wor94,har99,can99}) that found a correlation between
the soft X-ray emission measured with {\it ROSAT} and the radio core
emission of radio galaxies. It is therefore plausible that the X-ray
emission has an origin at the base of the radio jet, on
scales larger than any torus. This is not the only interpretation; for
example, \cite{don04} suggest that the soft X-ray emission may have an
origin in an accretion flow and that an obscuring torus is {\it
absent}. 

VLBI observations of parsec-scale radio jets (e.g.,~\citealt{pea96,gio01}) show evidence
for relativistic bulk motion, often with Lorentz factors in excess of
5, which implies that the radio cores observed with the VLA are likely
to be affected by beaming. Indeed, although the intrinsic radio powers of these sources are
similar, the distribution of core prominences (defined as the ratio of 5-GHz VLA core to 178-MHz total flux
densities) spans over four orders of magnitude. \cite{har00} demonstrated that the expected
distribution of core prominences of a randomly orientated population
of radio jets with a single intrinsic core prominence and bulk Lorentz
factor 5 replicates the distribution of observed core prominences, and
is consistent with the hypothesis that the unresolved VLA cores are
strongly affected by relativistic beaming. Therefore, the existence
of the radio--X-ray flux and luminosity correlations suggests that the
X-ray emission is also affected by beaming. The simplest mechanism for this is that the X-ray and radio
emission have a common origin in the relativistic electron population
in a jet. If the X-ray emission of components of low intrinsic
absorption were instead associated with the accretion flow, then the
observed correlations between the radio and X-ray fluxes and
luminosities would not necessarily be expected. Indeed, although
\cite{mer03} found a positive correlation between the X-ray and radio
luminosities in a sample of both radio-quiet (but not silent) and radio-loud
sources, the scatter is as much as 5 orders of magnitude. In our
present sample, which selects only radio-loud sources, the scatter in
X-ray components likely to have a jet origin is significantly less
than the radio-quiet sources in the \cite{mer03} sample, whose
X-ray emission is dominated by an accretion flow.

We note that there is no distinction between these components
of X-ray emission associated with FRI- and FRII-type radio galaxies:
each of the FRII sources, although being dominated by heavily obscured
emission, has a component of X-ray emission with low intrinsic
absorption that lies on the same trendline as the FRI
components. In Figure~\ref{ch7_fri_frii_indist_lum}, we show a plot of
the 1-keV luminosity density against the 5-GHz radio luminosity
density for the FRI- and FRII-type sources, which illustrates this
point.

The observed range in photon indices is consistent with either a
synchrotron or inverse-Compton mechanism for the jet-related X-ray
emission. The dispersion in the photon indices could be
explained by a two-component synchrotron+SSC model in which varying
the beaming parameter causes one component to dominate over the other.

\subsection{Components with high intrinsic absorption: An accretion origin}
\label{ch7_rxhigh}

In this section, we consider those X-ray components with intrinsic
absorption greater than $5\times10^{22}$ atoms cm$^{-2}$.  The mean
photon index of these components is $1.76\pm0.02$ and is significantly
flatter than that of components with low intrinsic absorption
($\Gamma=1.88\pm0.02$). The observed range of photon indices in these
components of X-ray emission is consistent with that measured in a
sample of ASCA observations of broad-line radio galaxies
(\citealt{sam99}). It is
evident from Figures~\ref{ch7_rxlumflux}a and~\ref{ch7_rxlumflux}b that the
unabsorbed X-ray luminosity and flux densities of these components lie
approximately 2 orders of magnitude above those components with
intrinsic absorption less than $5\times10^{22}$ atoms
cm$^{-2}$. Intrinsic absorption in excess of $10^{23}$ atoms cm$^{-2}$
is unlikely to be simply due to dust in the host galaxy, as this would
imply that very strong ($A_{\rm V}=50$) dust lanes and/or molecular clouds would have to be placed
fortuitously in each of these objects to obscure the nucleus. Instead,
we argue that the absorption is consistent with an origin in a dusty structure surrounding the
nucleus, such as the putative torus. From Figure~\ref{ch7_rxlumflux}b we
note that the correlation between the radio and X-ray flux densities
of these components appears weaker than those with $N_{\rm H} <
5\times10^{22}$ atoms cm$^{-2}$ that were interpreted to have a
physical origin in the form of a beamed radio jet. Indeed, analysis
with the Kendall $\tau$-test shows that the significance of this
correlation is 91.7\%. However, it is only by sampling a larger number
of FRII-type sources that we can determine if this
correlation is in fact weaker.

While the chosen cutoff of $5\times10^{22}$ atoms cm$^{-2}$ was in a sense arbitrary, an obvious physical difference in the emission processes of the two ``bands'' of X-ray emission manifests itself: those sources dominated by a component of emission with high intrinsic absorption all have fluorescent K$\alpha$ line-emission from neutral or near-neutral states of iron, whereas those with low intrinsic absorption do not. The line parameters are consistent with an origin in cold material, either from the outer regions of an accretion disk, or from a region still further out, possibly in the form of a torus-like structure that surrounds the nuclear emission.

\medskip

In summary, for components with high intrinsic absorption ($N_{\rm H} > 10^{23}$ atoms cm$^{-2}$), we have shown that:

\begin{enumerate}
\item Their unabsorbed X-ray flux and luminosity densities lie above those that are likely to have an origin in the form of a jet
\item All are associated with Fe K$\alpha$ lines
\end{enumerate}

The most probable interpretation for the X-ray emission of these
components, therefore, is that they are dominated by an
accretion flow and are surrounded by a dusty circumnuclear structure,
plausibly in the form of a torus. All narrow-line FRII-type sources show these features. This distinguishes them from
FRI-type sources, whose X-radiation is dominated by unabsorbed
emission, likely to be related to the jet (with the exception of
Cen~A). The one exception to the observed properties of components
with $N_{\rm H} > 5\times10^{22}$ atoms cm$^{-2}$ is 3C~388. Under the
assumption that its nuclear emission is obscured by a column of
$10^{23}$ atoms cm$^{-2}$, its unabsorbed X-ray flux and luminosity
densities lie below those of the other sources discussed
here. However, 3C~388 is the only low-excitation FRII-type radio
galaxy in the sample which, as postulated by \cite{har04a}, may imply
that it is an intrinsically low-jet-power (i.e., more FRI-like)
source, with its high 178-MHz radio luminosity due to a rich
surrounding environment (c.f., \citealt{bar96}). Our measurement of a relatively low X-ray core luminosity for this source is consistent with this argument.

The broad-line radio galaxy 3C~390.3 also shows the same X-ray
emission properties as those sources with high intrinsic absorption,
in that its X-ray flux and luminosity densities also lie above those
sources with low intrinsic absorption. The detection of broad optical
emission lines in this source implies, in the context of unified AGN
schemes, that it is oriented to the observer such that its accretion
system is viewed directly, leading to its high luminosity but low
intrinsic absorption. Other pieces of physical evidence support this
interpretation, such as the detection of rapid variability and Fe
K$\alpha$ line-emission (\citealt{ind94,lei97,gli03b}), although the
fluorescent iron line is only detected marginally  in our limited
signal-to-noise \Ch observation.

In order to consider the possible structure of the accretion flow in
the components with high intrinsic $N_{\rm H}$, we compared their X-ray
and Eddington luminosities. Out of these high-$N_{\rm H}$ sources, it
is only for Centaurus~A (\citealt{mar01})
and Cygnus A (\citealt{tad03}) that estimates of the black hole mass
from dynamical motions of stellar kinematics are available. The black
hole masses of the other sources, where available, are taken from
\cite{bet03} and \cite{marc04}, who assume that they lie on the
previously established correlation between the mass of the black hole
and the host bulge magnitude (e.g.,~\citealt{fer00}). As there exists
considerable scatter in this relation, these black hole estimates are
somewhat uncertain. Taking the black hole masses at face value, we
tabulate, for each source, its 0.5--10 keV unabsorbed X-ray
luminosity, Eddington luminosity, and the ratio of these two
quantities ($\eta_{\rm X,Edd}$). The results are shown in
Table~\ref{ch7_eddingtonefficiencies}.

Under the assumption that the sources accrete at the Eddington limit, from Table~\ref{ch7_eddingtonefficiencies}, it can be seen that $\eta_{\rm X,Edd}$ for these sources typically ranges from $\sim$10$^{-3}$ to $\sim$10$^{-2}$, with the value for Cen~A somewhat lower at $\sim$10$^{-5}$. However, it should be noted that the efficiency is based only on the 0.5--10 keV X-ray luminosity of the source, {\it not} its bolometric luminosity, which could typically be a factor of 3 to 10 times higher (\citealt{elv94}). Indeed, for Cygnus A (3C~405), \cite{tad93} estimate a bolometric luminosity of between $5\times10^{45}$ and $2\times10^{46}$ ergs s$^{-1}$, placing its efficiency at $\sim$5$\times10^{-2}$.

Although the efficiency values are uncertain, it seems that the accretion flows in these sources tend to have relatively high $\eta_{\rm X,Edd}$, which would suggest a significant contribution to the emission from a standard, geometrically thin, optically thick accretion disk (e.g.,~\citealt{sha73}), rather than a radiatively inefficient ADAF-type accretion flow (e.g.,~\citealt{nar95}). The efficiency of Cen~A is lower than those of the other components studied here, which may imply that its accretion flow takes the hybrid form of a radiatively inefficient optically thin inner component surrounded by a standard thin disk, as discussed more fully by \cite{evans05}.

\subsection{Optical constraints}
\label{ch7_optical}

In Figure~\ref{ch7_oxlum} we plot the unresolved red optical
luminosity densities (mostly measured using the F702W filter of the
{\it HST}'s WFPC2 instrument) and 1-keV X-ray luminosity densities of the
components of emission associated with each source, where {\it HST} data exist. As in
Section~\ref{ch7_rx}, we separate the X-ray components into those with
intrinsic absorption less than and greater than $5\times10^{22}$ atoms
cm$^{-2}$. Again, it can be seen that a good correlation exists
between the optical and X-ray luminosities and fluxes of those
components with intrinsic absorption less than $5\times10^{22}$ atoms
cm$^{-2}$, as found previously by, e.g., \cite{har00}. In a similar
manner to the radio and X-ray plots, components with intrinsic
absorption greater than $5\times10^{22}$ atoms cm$^{-2}$, together
with 3C~390.3, produce more X-ray emission than those with intrinsic absorption below this value.

The {\it HST} data allow a test of the postulate that the intrinsic
absorption of components with $N_{\rm H} < 5\times10^{22}$ atoms cm$^{-2}$ is associated with dust in the host
galaxy. {\it HST} observations of low-redshift FRI-type radio galaxies
in the 3C catalogue (\citealt{chi99}) reveal a variety of structures,
including dust lanes and a series of dusty disks oriented at a large
range of angles to the observer. Aside from Centaurus A, the two
sources with the highest intrinsic absorption below the imposed
$5\times10^{22}$ atoms cm$^{-2}$ cutoff, 3C~83.1B and 3C~296, are
associated with host galaxies with circumnuclear disks at high
inclinations to the observer. Figure~\ref{ch7_oxlum} shows that these two sources also have the highest deficit of optical emission
with respect to X-ray emission, which
indeed suggests that they are most affected by absorption.

\section{CONSTRAINTS ON A TORUS IN FRI-TYPE RADIO GALAXIES}
\label{ch7_fri_hidden}

If, as seems likely, the X-ray emission of the nuclei of FRI-type galaxies is dominated by the base of a relativistic jet that occurs on scales larger than that of any putative torus, one cannot determine directly the presence or absence of the torus, weakening the claim by \cite{don04} that the torus is absent in these sources. However, inferences can be made about its presence, and in this section, we test the hypothesis that a torus is present in FRI-type nuclei.

Let us assume that, in addition to the dominant jet component of X-ray
emission, there exists a `hidden' component of accretion-related
emission of photon index 1.7 obscured by a torus of intrinsic
absorption $10^{23}$ atoms cm$^{-2}$. This choice of obscuration and
photon index is consistent with that measured in Centaurus~A and the
heavily absorbed FRII-type radio galaxies discussed in
Section~\ref{ch7_rxhigh}, and the photon index is close to the value
measured for a sample of type 2 Seyfert galaxies observed with {\it
ASCA} (\citealt{tur97b}). For each `hidden' component, we then
consider the 90\%-confidence upper limits to its 0.5--10 keV
luminosity. As an example, consider 3C~264. Its nuclear spectrum is
modeled by a single, unabsorbed power law. Adding a component of
heavily absorbed emission to the model fit, refitting the spectra, and
determining the 90\%-confidence errors, yields an upper limit to the
0.5--10 keV luminosity of $3.3\times10^{40}$ ergs s$^{-1}$ for this
component. In Table~\ref{ch7_freezepow1_tab}, we repeat this exercise
for all FRI-type nuclei with low intrinsic absorption ($N_{\rm H} <
5\times10^{22}$ atoms cm$^{-2}$), and also show the black hole mass,
Eddington luminosity, and upper limits on $\eta_{\rm X,Edd}$ of each source. The black
hole masses quoted are taken from the black hole mass--host galaxy
magnitude correlation described in Section~\ref{ch7_rxhigh}, although
kinematic mass estimates are available for 3C~272.1 (M84) (\citealt{bow98}), 3C~274
(M87) (\citealt{ford94}), and NGC 6251 (\citealt{fer99}).

None of the values of the primary (detected) power-law
photon index of the unobscured component varies significantly with the
addition of this hidden component. We find that, under the assumption
of obscuration by a column of $10^{23}$ atoms cm$^{-2}$, the upper
limits to any hidden, accretion-related luminosities are all in the
range $10^{39}$--$10^{41}$ ergs s$^{-1}$. The highest hidden
luminosities cannot exceed that of the heavily absorbed (and
likely accretion-related) component measured in the FRI-type radio
galaxy Centaurus~A ($\sim$5$\times10^{41}$ ergs s$^{-1}$), but some
have maximum possible luminosities that are one to two orders of magnitude lower than
this. In addition, {\it all} have upper limits to the 0.5--10 keV
luminosities and Eddington efficiencies several orders of magnitude
less than those of the accretion-related components in the FRII
sources, whose mean luminosity is $\sim4\times10^{43}$ ergs
s$^{-1}$. (Again, we note that the low-excitation FRII-type radio
galaxy 3C~388 is an exception here). 

A similar calculation for an absorbing column of $10^{24}$ atoms
cm$^{-2}$ (Table~\ref{ch7_freezepow1_tab}) still does not permit
FRI-type sources to possess hidden accretion components with
efficiencies $\eta_{\rm X,Edd}$ comparable with those of the FRII-type
sources. This suggests that the accretion flows of FRI-type sources radiate at lower efficiency than do FRII-type sources, and may for example take the form of optically thin ADAFs (e.g.,~\citealt{nar95,esi97}).

\section{A NUCLEAR FANAROFF-RILEY DICHOTOMY?}
\label{ch7_dichotomy}

We have considered the correlations between the flux and luminosity
densities of the X-ray, radio, and optical components of nuclear
emission and argued that the X-ray emission of FRI-type radio-galaxy
nuclei is most likely dominated by emission from a parsec-scale jet,
with little or no intrinsic absorption. By contrast, the emission of
FRII-type radio-galaxy nuclei is dominated by an accretion flow and is
heavily absorbed (with the exception of the broad-line radio galaxy
3C~390.3) and therefore must be surrounded by a dusty structure, such
as the putative torus. In addition, each heavily absorbed component
has an accompanying component of X-ray emission of intrinsic
absorption less than $5\times10^{22}$ atoms cm$^{-2}$, the detections
or upper limits to the flux and luminosity densities of which all lie
in the region occupied by those sources likely to have a jet-related
origin (see Figure~\ref{ch7_fri_frii_indist_lum}).

Our results imply that, for
the FRII-type radio galaxies at least, both jet- and
accretion-related components of X-ray emission are present, which is
consistent with unified models of AGN. Further, we have shown that
the data do not exclude the presence of heavily obscured,
accretion-related emission in FRI-type radio galaxies, but that it is
of lower luminosity than in FRII-type radio galaxies if there is a
similar level of obscuration.

It is clear that there tends to be a dichotomy in the observed
properties of the X-ray nuclei of FRI- and FRII-type radio
galaxies. But why should this dichotomy occur? Is it due to 
intrinsic differences in the properties
of FRI and FRII radio galaxies, or simply due to the relative
contributions of jet- and accretion-related emission varying with the
total power of the source? In what follows, we discuss alternative
models to explain the observed differences in the X-ray emission
characteristics of FRI- and FRII-type sources.

\subsection{Model 1: FRI-type galaxies have tori of higher intrinsic absorption than FRII-type radio galaxies}
\label{ch7_model1}

In this model, a luminous accretion flow in an FRI-type radio galaxy
(of luminosity consistent with those measured in the FRII-type radio
galaxies) is surrounded by a torus of higher intrinsic absorption than
the FRII-type radio galaxies. Such a model gives rise to a reduced
contribution from the accretion system in terms of observed X-ray
flux, such that the jet dominates the X-ray emission. In
Table~\ref{ch7_freezepow1_tab}, we showed that even if a Compton-thick
($N_{\rm H} \sim 10^{24}$ atoms cm$^{-2}$) torus were to obscure the
accretion-related emission of jet-dominated FRI-type sources, the
luminosities and efficiencies are still somewhat lower than those of
the FRII-type sources (see
Table~\ref{ch7_eddingtonefficiencies}). Thus it seems that if a torus
obscures FRI-type accretion flows, the luminosities of FRII-type
accretion flows are still higher, unless the intrinsic absorption of
the torus in FRI-type sources is extreme ($N_{\rm H} > 10^{24}$ atoms cm$^{-2}$).

Evidence against this model comes from constraints from infrared data. If FRI-type radio galaxies do indeed harbor luminous accretion flows surrounded by very dusty structures, one would expect to detect significant infrared emission in the centers of these sources. Such a scenario may be ruled out from {\it ISO} observations of a sample of 3CR FRI- and FRII-type radio galaxies (\citealt{mul04,haas04}). None of the FRI sources exhibits any high MIR or FIR dust luminosity, which would be expected for an intrinsically powerful, but highly obscured AGN (c.f., the FRII sources, in which strong MIR and FIR emission is observed). 

Based on the above arguments, it seems unlikely that the accretion flows of FRI-type sources are obscured by tori of higher intrinsic absorption than in FRII-type sources. This then implies that FRI-type accretion flows are less luminous than their FRII-type counterparts, either due to a lower radiative efficiency or lower accretion rate. In the immediate future, {\it Spitzer} observations of all 36 3CRR sources at $z < 0.1$ (Birkinshaw et al., in prep.) will test this hypothesis further.

\subsection{Model 2: The relative contribution of accretion-related and jet-related emission varies smoothly as a function of total AGN power}
\label{ch7_model2}

In this model, the fraction of jet- and accretion-related emission
varies with AGN power in such a way that, at higher AGN powers,
accretion-related emission comes to dominate over jet-related
emission. In this scenario, an intrinsic dichotomy between the properties
of FRI- and FRII-type radio-galaxy nuclei need not exist, and the
observed properties would simply be due to a smooth scaling relation
with AGN power. The 178-MHz
luminosity density of a source is an approximate measure of its AGN
power since (1) it correlates reasonably well with jet power
(\citealt{fr74}) and (2) jet power correlates reasonably well with the
narrow-line luminosity of AGN (\citealt{raw91}). 

In order to test this model, in
Figure~\ref{ch7_178_corr} we plot the 1-keV X-ray accretion-flow
luminosity density (either modeled from the spectra in the case of the
FRII-type sources and Cen~A, or upper limits in the case of the
FRI-type sources) against the 178-MHz luminosity density. Although there is
considerable scatter, it is plausible
that there exists an underlying correlation between the 178-MHz luminosity
densities and 1-keV luminosity densities of the accretion-related
components, which may support this model.

This model can explain the observed differences in the X-ray emission
characteristics of FRI- and FRII-type sources {\it without} having to
invoke an intrinsic dichotomy in the nuclear properties of the two
populations. Nevertheless, this model leaves unexplained the fact that
the transition between the X-ray emission of a source being
jet-dominated and accretion-dominated should occur so close to the
FRI/FRII boundary, rather than there being a significant population of
jet-dominated FRII-type sources or, conversely, accretion-dominated
FRI-type sources. However, a possible resolution to this problem is in
the model proposed by \cite{fal95}, in which the opening angle of the
torus is proportional to the power of the accretion flow. At low
accretion-flow power, the opening angle of the torus is such that
material from it becomes stripped and entrained into the jet flow,
eventually causing the observed deceleration of the jet on kiloparsec
scales, leading to an FRI-type large-scale morphology. As soon as the
torus opening angle becomes greater than the opening angle of the jet,
the entrainment is not as severe, allowing the jet to remain highly
relativistic and produce an FRII-type source.

Another potential problem with this model is that at the FRI/FRII boundary of $\sim$10$^{25}$ W Hz$^{-1}$ sr$^{-1}$, the accretion-related luminosity of the FRII-type source 3C~98 is 1.5 orders of magnitude greater than the upper limits on that in the jet-dominated FRI-type sources 3C~338 and 3C~465. In other words, it is impossible to `hide' an accretion flow of the luminosity of 3C~98 in these two jet-dominated sources of comparable 178-MHz power, unless the intrinsic absorption is extreme. This may at first suggest that the luminosity of the accretion-related emission does {\it not} scale smoothly with the 178-MHz luminosity density. However, we note that the model of \cite{fal95} may provide an adequate explanation of the above effect, and that only these 3 of the 8 3CRR sources that span the boundary have been observed with {\it Chandra} or {\it XMM-Newton}. It is important to test this model by observing the remaining 5 3CRR sources.

\subsection{Model 3: An intrinsic dichotomy exists in the accretion-flow structures of FRI- and FRII-type sources}
\label{ch7_model3}

It has previously been proposed (e.g.,~\citealt{rey96,don04}) that there exists a fundamentally different accretion {\it mode} in FRI- and FRII-type sources, such that the accretion-flow luminosities and radiative efficiencies of FRI-type radio galaxies are systematically lower than those of FRII-type radio galaxies, consistent with our observations.

The most widely discussed interpretation in this context is one in which the fractional mass accretion rate $\dot{m}=\dot{M}/M$ governs the contribution to the emission from a radiatively inefficient optically thin advection-dominated accretion flow (ADAF). \cite{esi97} argued for black-hole X-ray binaries that there exists a critical fractional mass accretion rate $\dot{m}_{\rm crit}$ below which gas is unable to cool efficiently, such that the accretion-flow energy is advected into the black hole, forming an ADAF. Above $\dot{m}_{\rm crit}$, the accretion flow makes the transition to being dominated by a standard, radiatively efficient, geometrically thin, optically thick disk (e.g.,~\citealt{sha73}), with a step increase in the total accretion-flow luminosity. By analogy, in radio galaxies, the accretion flow of FRI-type radio galaxies may take the form of a radiatively inefficient ADAF-type model, whereas in FRII-type galaxies it is more likely to form a Shakura-Sunyaev disk. 

This model resolves the two main issues of Model 2. Firstly, it no longer requires there to be a coincidence that the transition from a source being jet-dominated to accretion-dominated occurs at the FRI/FRII divide, since this is a consequence of the different accretion-flow modes in FRI- and FRII-type sources. Secondly, it explains why the luminosity of the accretion flow of 3C~98 is significantly higher than those of the two jet-dominated FRI-type sources that lie at comparable 178-MHz luminosity: the accretion-flow luminosity of 3C~98 has a significant contribution from a Shakura-Sunyaev accretion disk, whereas those of 3C~338 and 3C~465 are likely to be dominated by a radiatively inefficient, optically thin accretion flow. Again, however, it is important to test this model by observing the remaining sources that populate the FRI/FRII break in 178-MHz radio luminosity.

The major uncertainty is how a dichotomy in the subparsec-scale accretion-flow mode could influence
the deceleration of jets into FRI- and FRII-type structures, which is observed to
occur on kiloparsec scales (e.g.,~\citealt{bic95,gkw00,mul04}). For
this model to remain viable, we
speculate that, although differing accretion-flow modes may exist, the
jet-production mechanism must be the same in the nuclei of FRI- and FRII-type sources.

One notable exception to the above models is the nuclear emission of
Centaurus~A. As the only FRI-type radio galaxy whose X-ray emission is
heavily obscured and dominated by an accretion flow, its nuclear X-ray
emission is more similar to that of the FRII-type sources. One
possible resolution of this problem is that the recent merger that has
occurred in Cen~A may have provided additional material to accrete
onto the supermassive black hole, triggering heightened nuclear
activity and a new phase of radio activity, causing the supersonic
re-inflation of the lobes (\citealt{kra03}).

\section{CONCLUSIONS}
We have presented results from a \Ch and \XMM spectral analysis of the nuclei of a sample of the nuclei of 22 low-redshift ($z < 0.1$) radio galaxies.  We find that:

\begin{enumerate}

\item The nuclear X-ray spectra of FRI-type sources are unabsorbed, or absorbed simply by gas related to the known kpc-scale dusty
disks in the host galaxy. The strong observed correlations between
the X-ray, radio, and optical fluxes and luminosities imply that the emission has a common origin at the base of a
relativistically beamed parsec-scale jet.

\item The nuclear X-ray emission of
narrow-line FRII-type sources is dominated by heavily absorbed
components of emission with $N_{\rm H} > 10^{23}$ atoms cm$^{-2}$, and
is accompanied by emission from neutral fluorescent Fe K$\alpha$
lines. We argue that this absorbed emission is likely to originate in
an accretion flow and be surrounded by a structure such as the
putative torus. We also find that the nuclear X-ray spectrum of every
FRII galaxy has a corresponding component of soft X-ray emission,
which is consistent with having a jet-related origin.

\item If the (jet-dominated) X-ray
emission of FRI-type sources occurs on scales larger than the torus,
it is impossible to test for the presence of a torus using the X-ray data, but important constraints can still be made. We
estimate the maximum level of a `hidden', accretion-related component of
emission that could be obscured by an adopted column of $10^{23}$
atoms cm$^{-2}$ to be in the range $10^{39}$--$10^{41}$ ergs
s$^{-1}$. The X-ray data do not exclude the presence of a torus,
but the luminosity of the accretion flow it obscures is
significantly less than in FRII-type sources unless there is more
obscuring matter in the FRI-type sources, which seems unlikely based
on infrared constraints.

\item Any `hidden' accretion flows in jet-dominated FRI-type sources are
likely to be significantly sub-Eddington in nature. This implies that their accretion flows are mass-starved, and/or radiate at a low efficiency.

\item The accretion-flow luminosities of FRII-type sources are typically several orders of magnitude higher than those of FRI-type sources. The ratio of X-ray to Eddington luminosities, $\eta_{\rm X,Edd}$, is $\sim$$10^{-3}$--$10^{-2}$, while the ratio of bolometric to Eddington luminosities is still higher. This implies that the accretion flows of FRII-type sources tend to be fed at a high rate, and/or possess significant contributions from high-radiative-efficiency flows, plausibly in the form of a standard, geometrically thin, optically thick disk.

\item Two models may account for the observed differences in the
nuclear properties of FRI- and FRII-type sources, although neither is
without problems. One model, in which the relative contribution of the
accretion-related and jet-related emission varies smoothly as a
function of total AGN power, can successfully account for the observed
X-ray emission characteristics of these sources. However, it is then
difficult to understand why the transition between a source being jet-dominated
and accretion-dominated occurs at the FRI/FRII boundary. Alternatively, there is a real dichotomy in the accretion-flow
modes of FRI- and FRII-type sources. We note that the manner in which the accretion-flow mode might
then affect the large-scale (FRI versus FRII) characteristics of radio
galaxies remains poorly understood.

\end{enumerate}

\acknowledgements

We are grateful for support for this work from PPARC (a Studentship
for D.A.E. and research grant for D.M.W.), the Royal Society (Research
Fellowship for M.J.H.), and NASA (contracts NAS8-38248 and NAS8-39073
with the Smithsonian Astrophysical Observatory). We thank Oliver
Shorttle for providing the results from his {\it HST} analyses of
these sources, and Elena Belsole for useful discussions. We are
grateful to the anonymous referee for useful comments.

\appendix
\section{NOTES ON INDIVIDUAL SOURCES}

\subsection{3C~31}

We initially attempted to model the source spectrum with a single,
unabsorbed, power law. However, the fit was poor ($\chi^2 = 49.3$ for
27 dof), with strong residuals at $\sim$1 keV, suggesting that a
contribution from thermal emission is necessary. The fit was
substantially improved ($\Delta\chi^2=39.7$ for two additional
parameters) by adding a thermal ({\sc{Apec}}) model of temperature
$kT=0.68^{+0.08}_{-0.07}$ keV, abundance 0.3 of solar, and
normalization ($3.39^{+0.80}_{-0.85})\times10^{-5}$. No further
significant improvement was obtained by including intrinsic absorption
in the model fit. The 1-keV flux density and
spectral index are consistent with previously
published \Ch results (\citealt{har02b,don04}). The 0.5--5 keV radial
surface-brightness shows a deficit of counts approaching a factor of 2
compared with those measured
spectrally. As discussed by \cite{har02b}, the discrepancy may be resolved by postulating that there exists an additional, unresolved dense component of thermal emission that lies close to the core.

\subsection{3C~33}

A variety of simple models were fitted to the nuclear spectrum but all
yielded poor results. The first acceptable fit ($\chi^2 = 49.0$ for 39
dof) was obtained with a model consisting of a heavily absorbed power
law [$N_{\rm H} = (3.9^{+0.7}_{-0.6}) \times10^{23}$ atoms cm$^{-2}$],
a $\sim$6.4 keV Gaussian Fe K$\alpha$ emission line of equivalent
width $320^{+340}_{-160}$ eV, and a second,
unabsorbed power law. When adding intrinsic absorption to the second
power-law component, the best fitting column density tended to 0. No
significant improvement in the fit ($\Delta\chi^2=8.21$ for three
additional parameters) was obtained by adding a component of thermal
emission (the probability of achieving a larger $F$ by chance is
8.2\%). The photon indices of both power laws were fixed at 1.7, owing
to the large number of free parameters already present in the
model. The 1-keV flux densities are largely insensitive to this choice, however.

\subsection{3C~66B}

A model consisting of a single, unabsorbed, power law provided a good
fit to the data ($\chi^{2} = 40.3$ for 36 dof). The best-fitting power
law photon index is $\Gamma = 2.25 \pm 0.11$. Adding thermal emission
of temperature $0.41^{+0.23}_{-0.10}$ keV, abundance half of solar,
and normalization $(2.67^{+1.76}_{-1.60})\times10^{-5}$ to the spectral model
significantly improved the fit ($\chi^{2} = 29.1$ for 34 dof). In this
case the photon index is 2.03 $\pm$ 0.18. A model consisting of an
absorbed power law and thermal emission failed to improve the fit
($\Delta\chi^{2}=2.7$ for 1 additional parameter, with a probability
of achieving a greater $F$ by chance of 7.6\%). We compared the
results of our spectral analysis with previously published work
(\citealt{har01,don04}), and found the 1-keV power-law flux density to be consistent.

\subsection{3C~83.1B (NGC 1265)}

We attempted to fit several models to the nuclear spectrum of 3C~83.1B,
but found that the only one that gave an acceptable fit ($\chi^{2} =
9.45$ for 12 dof) consisted of the sum of an absorbed power law and thermal
emission of abundance 0.3 of solar. The best fitting spectral parameters for this model are
$N_{\rm H} = (3.2^{+0.8}_{-0.7}) \times 10^{22}$ atoms cm$^{-2}$,
$\Gamma = 2.00^{+0.27}_{-0.20}$. The temperature of the thermal
component is $0.58^{+0.15}_{-0.14}$ keV, with normalization $(9.69^{+2.95}_{-2.74})\times10^{-6}$. These parameters are
consistent with those measured by \cite{sun05}, who performed the
original observation.

\subsection{3C~84 (NGC 1275)}

A model fit consisting of power law plus a single thermal component
failed to provide an adequate fit to the spectrum ($\chi^2 = 1322.8$
for 1155 dof). An acceptable fit ($\chi^2 = 1165.3$ for 1153 dof) was
achieved with the combination of a power law and {\it two} thermal
components, one of temperature $0.77 \pm 0.04$ keV, solar
abundance, and normalization $(8.33^{+1.00}_{-1.33})\times10^{-4}$;
the other of temperature $2.74 \pm 0.10$ keV, abundance half of solar,
and normalization $(9.72^{+0.50}_{-0.72})\times10^{-3}$. However, a third thermal component, fitted by
\cite{don04}, is not found here. Instead, an additional significant
improvement in the fit ($\Delta\chi^2=9.35$ for two additional
parameters) was achieved by the addition of a Gaussian emission line
with a centroid energy $6.39^{+0.08}_{-0.09}$ keV and frozen
(unresolved) linewidth of 10 eV and equivalent width $26^{+46}_{-22}$ eV. We compared our results with
previously published work (\citealt{don04}), and find the 0.3--8 keV unabsorbed luminosity of the power law to be approximately consistent.

\subsection{3C~98}
\label{3crr_3c98}

A model fit consisting of a heavily absorbed power law [$N_{\rm H} =
(1.1 \pm 0.2) \times10^{23}$ atoms cm$^{-2}$; $\Gamma = 1.56 \pm
0.26$] and thermal emission [$kT = 0.98 \pm 0.12$ keV] provided a good
fit to the data ($\chi^2 = 40.7$ for 42 dof). However, an adequate fit
($\chi^2 = 47.1$ for 42 dof) was also achieved with the sum of two
power laws, one heavily absorbed [$N_{\rm H} = (1.5 \pm 0.2)
\times10^{23}$ atoms cm$^{-2}$; $\Gamma = 1.87 \pm 0.32$], and the
other with no absorption and a frozen photon index of 2. Both fits
were improved when an unresolved Gaussian Fe K$\alpha$ line of linewidth frozen at 10 eV was added; the improvements in the fits were $\Delta\chi^2=8.0$ and $\Delta\chi^2=7.4$, respectively, for two additional parameters. A model consisting of a heavily absorbed power law, an unabsorbed soft power law, a Gaussian Fe K$\alpha$ line and thermal emission, did not significantly improve the fit over the previous two models ($\chi^2 = 30.7$ for 39 dof). 

{\it ROSAT} observations of 3C~98 (\citealt{har99}) show extended
X-ray emission on scales of tens of arcseconds, in addition to an
unresolved core. It is therefore likely that the most appropriate
spectral model for this source is the one consisting of a heavily
absorbed power law [$N_{\rm H}=(1.2^{0.3}_{-0.2})\times10^{23}$;
$\Gamma=1.68^{+0.23}_{-0.37}$], a Gaussian Fe K$\alpha$
[$E=6.37\pm0.10$ keV; equivalent width
$240^{+250}_{-170}$ eV], and thermal emission [$kT=0.98\pm0.12$ keV;
abundance 0.3 of solar; normalization $(6.21\pm1.33)\times10^{-5}$]. We note that the upper limit on the flux density of the
statistically insignificant unabsorbed power law (which might be
regarded as jet-related nuclear emission) is an interesting quantity. We compared the results of our \XMM spectral fitting with previously published work (\citealt{isobe05}). These authors measure a 2--10 keV unabsorbed luminosity of $(4.6^{+0.7}_{-0.6})\times10^{42}$ ergs s$^{-1}$, consistent with the value we measure of $(5.4^{+1.4}_{-2.6})\times10^{42}$ ergs s$^{-1}$. The photon indices and temperature of the thermal emission are also consistent.

\subsection{3C~264}
\label{3crr_3c264}

The nuclear spectrum is well modeled by a single, unabsorbed, power
law of photon index $2.34^{+0.07}_{-0.08}$. For this fit, we found the
value of $\chi^2$ to be 123.0 for 147 dof. No improvement to the fit was achieved with more complex
spectral models, such as an intrinsically absorbed power law or the
addition of thermal emission: in every case, the values of absorption
and thermal normalization tended to zero. We compared the results of
our spectral fitting to that from an \XMM observation
(\citealt{don04}). The power-law photon indices are approximately
consistent ($\Gamma=2.34^{+0.07}_{-0.08}$ for \Ch and
$\Gamma=2.48\pm0.04$ for {\it XMM-Newton}), as are the integrated
X-ray luminosities.

\subsection{3C~272.1 (M84)}

A single power law provided an acceptable fit to the nuclear spectrum
($\chi^2 = 30.3$ for 25 dof). However, the fit was significantly
improved ($\Delta\chi^2=19$ for one additional parameter), with the
inclusion of relatively mild intrinsic absorption [$N_{\rm H} =
(1.9^{+0.9}_{-0.7}) \times10^{21}$ atoms cm$^{-2}$] at the redshift of
3C~272.1. No statistically significant improvement in the fit was
achieved with the inclusion of a thermal {\sc Apec} model, which
implies that the local background subtraction has accounted for most
of the extended thermal emission. We compared the results of our
spectral fitting with previously published \Ch work (e.g.,
\citealt{harris02}), and found the power-law photon index,
normalization and intrinsic absorption to be consistent. Using the same
\Ch data, \cite{don04} determined the best-fitting spectral model to
be the sum of an absorbed power law and thermal emission. The
detection of thermal emission by \cite{don04} is likely due to the selection of an
off-source region from which to extract the background spectrum. The
intrinsic absorption and photon index we measure agree with the
analysis by \cite{don04}, and the integrated power-law luminosity is
approximately consistent.

\subsection{3C~274 (M87)}

A model fit to the spectrum consisting of a single, unabsorbed, power
law provided an acceptable fit to the data ($\chi^2 = 97.8$ for 100
dof). However, this fit was significantly improved
($\Delta\chi^2=13.6$ for two additional parameters) with the addition
of a thermal component, characterized by an {\sc Apec} model of
temperature $0.75^{+0.17}_{-0.14}$ keV, abundance 0.3 of solar, and
normalization $(1.11\pm0.49)\times10^{-4}$. Such a temperature in the
inner regions of M87 might not be unexpected, as shown from an \XMM
study of the radially dependent temperatures of the hot,
X-ray-emitting gas in this source (\citealt{boh01}). The power-law
photon index for this fit is 2.09 $\pm$ 0.06. For this model, $\chi^2
= 84.2$ for 98 dof, with the probability of achieving a greater $F$ by
chance 0.07\%. No further statistically significant improvements to
the fit were achieved with more complex spectral models. We compared
the results of our \Ch nuclear spectral analysis with those found by
{\cite{wil02}. The power-law photon indices and normalizations are
consistent, although \cite{wil02} found evidence for slight additional absorption ($N_{\rm H}=3.5^{+1.5}_{-1.4}\times10^{20}$ atoms cm$^{-2}$) at the redshift of M87.

\subsection{3C~296}

Several simple one-component models were fitted to the nuclear
spectrum, but we found that the first acceptable fit was achieved with
a model consisting of an absorbed [$N_{\rm H} = (1.5 \pm 0.9)
\times10^{22}$ atoms cm$^{-2}$] power law of photon index
$1.77^{+0.60}_{-0.52}$, accompanied by thermal emission [$kT =
0.8^{+0.3}_{-0.5}$ keV; abundance 0.3 of solar; normalization $(1.16^{+0.42}_{-0.45})\times10^{-5}$]. The fit was good: $\chi^2 = 9.3$ for 21
dof. However, an acceptable fit was also achieved with the combination
of an absorbed power law with $N_{\rm H} = (2.1^{+1.5}_{-1.7})
\times10^{22}$ atoms cm$^{-2}$ and photon index $\Gamma =
1.9^{+0.8}_{-0.7}$, and a {\it second} unabsorbed power law with a
photon index frozen at 2. A subsequent radial-profile analysis shows
that the number of extended counts measured spectrally and spatially
agree for the first model, and so we adopt this model. We performed a comparison between our spectral analysis and previously published \Ch results (\citealt{har05}). The measured values of the intrinsic absorption, power law photon index, and 1-keV flux densities are consistent, as are the parameters of the thermal component.

\subsection{3C~321}

Single-component models provided poor fits to the spectrum, with
noticeable residuals at high and low energies, together with a
residual at $\sim$6 keV suggesting the presence of Fe K$\alpha$
line-emission. The first moderately acceptable fit ($\chi^{2} = 19.2$
for 9 dof) was achieved with an absorbed [$N_{\rm H} = (8.7 \pm
5.7)\times10^{21}$ atoms cm$^{-2}$] power law, a Gaussian Fe K$\alpha$
emission line, and a thermal component of temperature 0.57 $\pm$ 0.04
keV. However, positive residuals at energies $\gappeq$4 keV were
noticeable, suggesting the presence of heavily absorbed emission. A model consisting of a heavily absorbed power law of
photon index frozen at 1.7, a strong Gaussian Fe K$\alpha$ line of
equivalent width $\sim$1 keV, a second
unabsorbed power law of photon index frozen at 2, and thermal emission
of temperature 0.49 $\pm$ 0.15 keV and normalization $(1.33^{+0.53}_{-0.34})\times10^{-6}$, provided the best fit to the data
($\chi^{2} = 9.8$ for 8 dof). Although the
photon indices are frozen at their canonical values [owing to the
relatively small number of bins (14) and large number of free
parameters], the parameter uncertainties are large.

\subsection{NGC 6109}

Single-component models of either thermal emission or a power law
provided an adequate fit to the nuclear spectrum ($\chi^{2} = 7.67$
for 8 dof, and $\chi^{2} = 7.46$ for 8 dof, respectively). However, a
significant improvement in the fit ($\chi^{2} = 1.50$ for 6 dof) was achieved with the combination
of a power law [$\Gamma=1.47\pm0.47$] and thermal emission
[$kT=0.63^{+0.14}_{-0.17}$ keV, abundance 0.3 of solar, normalization $(1.01^{+0.41}_{-0.42})\times10^{-6}$].

\subsection{3C~338}

Simple one-component models of either thermal emission or a power law
provided adequate fits to the data ($\chi^2 = 1.66$ and 4.47 for 3
dof, respectively). The best-fitting parameters for the power-law
model are $\Gamma = 2.37 \pm 0.81$ with a 1-keV normalization
$(5.2^{+2.1}_{-2.1}) \times 10^{-6}$ photons cm$^{-2}$ s$^{-1}$
keV$^{-1}$. A radial surface-brightness profile showed a clear excess
of counts over extended emission at distances $<1''$ from the core, and so we
adopt the best-fitting model for the nuclear spectrum to be a single,
unabsorbed, power law. We compared the results of my \Ch spectral
fitting to a previously published \Ch analysis of 3C~338 (\citealt{dimatt01}). These authors found an unabsorbed 1-keV flux density of $(7\pm2)\times10^{-15}$ ergs cm$^{-2}$ s$^{-1}$ keV$^{-1}$, which is consistent with the value of $(10.6\pm2.8)\times10^{-15}$ ergs cm$^{-2}$ s$^{-1}$ keV$^{-1}$ that we measure (errors here are 1$\sigma$ for one interesting parameter). 

\subsection{NGC 6251}

The \Ch observation of the nucleus of NGC 6251 has already been analyzed in detail by us
(\citealt{evans05}), and this paper should be consulted for a detailed description. Its nuclear spectrum is described
by an absorbed power law of $N_{\rm H} = 4.5 \times 10^{20}$ atoms
cm$^{-2}$ and photon index $\Gamma = 1.67 \pm 0.06$, and is mixed with
small-scale thermal emission of temperature $kT = 0.20 \pm 0.08$
keV, abundance 0.35 of solar, and normalization $(1.04^{+1.95}_{-0.48})\times10^{-4}$. Comparisons with other work are discussed by \cite{evans04}.

\subsection{3C~388}

An excess of counts at a position coincident with the nucleus is seen
at energies of 3--7 keV, and the
distribution of 3--7 keV counts is more point-like than an image
created from 0.5--1 keV counts. This suggests that the nucleus is
indeed detected, although there are relatively few counts. In
order to provide some constraints on the nuclear emission of 3C~388, we
adopted a spectral model consisting of an unabsorbed power law of
photon index frozen at 2.0, together with thermal emission
characterized by an {\sc Apec} model of temperature and abundance
frozen at 1.5 keV and 0.3 of solar, respectively. The thermal parameter
values are consistent with those found from an analysis of the
dependence of the hot-gas structure of 3C 388 as a function of the
distance from the nucleus (Kraft et al., submitted). The fit was good
($\chi^{2}=7.7$ for 9 dof), and the 1-keV unabsorbed flux density of
the power law we measure with \Ch is 6$\pm$5 nJy. As an additional
comparison with the other heavily absorbed FRII-type radio galaxies
that we analyze in this paper, we measured the 1-keV unabsorbed flux density of the
power law in a model fit consisting of an absorbed power law of
$N_{\rm H}=10^{23}$ atoms cm$^{-2}$ and photon index frozen at 2.0,
together with thermal emission. This value is 11 $\pm$ 6 nJy.

\subsection{3C~390.3}
\label{3crr_3c390.3}

A model fit to a single, unabsorbed, power law of photon index 1.78
$\pm$ 0.09 provided a good fit to the nuclear spectrum ($\chi^{2}
= 24.86$ for 54 dof). No improvement in the fit was achieved with the
inclusion of intrinsic absorption at the redshift of 3C~390.3: the
best-fitting value of $N_{\rm H}$ was zero, with a 90\% confidence
upper limit of $2.6\times10^{20}$ atoms cm$^{-2}$. Inspection of the
$\chi^{2}$ residuals of the best-fitting model showed slight positive
residuals at energies $\sim$6--7 keV, which suggests that Fe K$\alpha$
line-emission may be present, although its detection is not significant.

We measure the 2--10 keV unabsorbed luminosity of 3C~390.3 to be $\sim3 \times10^{44}$ ergs s$^{-1}$. Previous X-ray studies of 3C~390.3 show a range of 2--10 keV
luminosities for this source: an {\it ASCA} observation
(\citealt{sam99}) measured a photon index of 1.75 $\pm$ 0.02 and a
2--10 keV luminosity of $\sim1 \times10^{44}$ ergs s$^{-1}$, while
long-term monitoring of the source with RXTE (\citealt{gli03b})
measured a photon index of 1.72 $\pm$ 0.02, and showed that the flux
varies by a factor of over 2 on timescales of days.

\subsection{3C~449}

A model fit consisting of an unabsorbed power law of photon index
$2.14^{+0.25}_{-0.19}$ and a thermal ({\sc Apec}) model of temperature
$kT = 1.03 \pm 0.10$ provided a good fit to the spectrum ($\chi^{2} =
37.5$ for 31 dof). The temperature of the gas is consistent with that
measured by \cite{cro03}. However, the 1-keV (aperture corrected) flux
density of the nuclear power law measured with \XMM is a factor of
2--3 higher than that measured with {\it Chandra}. Although we note that variability
is a possible explanation for this effect, we decided to perform a
further investigation by extracting a spectrum from the \Ch data using
a source-centered circle of radius identical to that used for the \XMM
spectral extraction. A good fit was obtained to
the \Ch spectrum with a model consisting of a power law and thermal
emission, and the flux densities of the power law measured with \Ch
and \XMM were consistent, and are also consistent with previously
published \XMM data (\citealt{don04}). This highlights the importance of \Ch to our
investigation, as its ability to separate spatially nuclear emission
from unrelated (in this case thermal) emission is highly desirable.

\subsection{3C~452}
\label{3crr_3c452}

Many spectral models were fitted to the data, but the only acceptable
fit was found following \cite{isobe02} (who performed the original
observation). This model fit consisted of a heavily absorbed power law
[$N_{\rm H} = (5.7^{+0.9}_{-0.8}) \times10^{23}$ atoms cm$^{-2}$;
$\Gamma = 1.7$ (frozen)], a narrow 6.4-keV Fe K$\alpha$ Gaussian
emission line of equivalent width $160^{+180}_{-130}$ eV , reflection from a `slab' of neutral material (such as
the surface of an accretion disk or a torus), and thermal emission,
characterized by an {\sc Apec} model of temperature 0.63 $\pm$ 0.29
keV, abundance 0.4 of solar, and normalization $(2.84^{+20.06}_{-1.42})\times10^{-6}$. The fit was good: $\chi^{2} = 65.9$ for 69 dof, and the parameter values we found are
consistent with those measured by \cite{isobe02}. No significant
improvement to the fit was achieved with the addition of a second,
unabsorbed power law of photon index frozen at 2. However, we note
that the upper limit of the flux density of this unabsorbed component
(which might be regarded as jet-related nuclear emission) is an
interesting quantity.

\subsection{3C~465}

A model fit consisting of a power law provided a poor fit to the
nuclear spectrum, with strong residuals at $\lappeq$1 keV, clearly
suggesting the presence of thermal emission on small scales. A good
fit ($\chi^{2}=26.1$ for 24 dof) was achieved with the combination of
an unabsorbed power law and a thermal ({\sc Apec}) model. However, the
photon index of the power law ($\Gamma = 1.19^{+0.26}_{-0.29}$) was
surprisingly flat. Instead, a significant improvement in the fit
($\Delta\chi^{2} = 3.9$ for one additional parameter) was achieved
when the power law was allowed to have some intrinsic absorption. In
this case, we measure the absorption to be $(4.4^{+6.0}_{-3.9}) \times
10^{21}$ atoms cm$^{-2}$, the photon index to be
$1.86^{+0.72}_{-0.52}$, and the temperature of the thermal component
to be $0.70^{+0.07}_{-0.06}$ keV (with abundance 0.3 of solar, and
normalization $(4.51^{+0.90}_{-1.00})\times10^{-5}$). Models containing emission from a
second power law instead of the thermal component failed to provide adequate descriptions of the
spectrum. We compared the results of our analysis with previously
published work (\citealt{har05b}), and found all
the spectral parameters to be consistent.

\subsection{3C~403}

A model fit consisting of a heavily absorbed [$N_{\rm
H}=4.5^{+0.7}_{-0.6}\times10^{23}$ atoms cm$^{-2}$] power law of
photon index 1.76 $\pm$ 0.23, an unresolved Gaussian Fe K$\alpha$ line
and a second, unabsorbed, power law of photon index frozen at 2,
provided a good fit to the data ($\chi^{2}=31.1$ for 50 dof). The
addition of the Gaussian Fe K$\alpha$ line is significant at
99.99\% on an $F$-test. Thw equivalent width of this line is
$220^{+60}_{-150}$ eV. The addition of a thermal {\sc Apec} component
did not significantly improve the fit ($\chi^{2}=29.0$ for 48 dof,
with the probability of achieving a greater $F$ by chance of
18.4\%). We compared the results of the spectral fitting to previously
published \Ch data (\citealt{kra05}) and found all parameter values to
be consistent.

\subsection{3C~405 (Cygnus A)}

A good fit to the spectrum ($\chi^{2}=69.4$ for 84 dof) was achieved
with the combination of a heavily absorbed [$N_{\rm
H}=(1.7\pm0.3)\times10^{23}$ atoms cm$^{-2}$] power law of photon
index $\Gamma=1.6\pm0.5$, a Gaussian Fe K$\alpha$ line of equivalent
width $250^{+210}_{-170}$ eV, and a second,
unabsorbed, power law of photon index frozen at 2. The addition of a
thermal component to this fit did not significantly improve the fit
($\Delta\chi^{2}=1.7$ for two additional parameters), with probability
of achieving a greater $F$ by chance of 36.4\%. These best-fitting
parameters are consistent with previously published work by
\cite{you02}, who used the same \Ch data.

\subsection{Centaurus A}

The nucleus of Centaurus~A has been analyzed in detail by us
(\citealt{evans04}), and this work should be consulted for a detailed description. The nuclear spectrum is well described
by a heavily absorbed ($N_{\rm H}\sim10^{23}$ atoms cm$^{-2}$) power
law of photon index 1.7, accompanied by a narrow fluorescent Fe
K$\alpha$ emission line of equivalent width $60\pm15$ eV. In addition, \cite{evans04} find that a contribution from
a softer power law, related to the parsec-scale VLBI jet, is
necessary to model the nuclear continuum.

\begin{figure}
\begin{center}
\epsscale{0.7}
\plotone{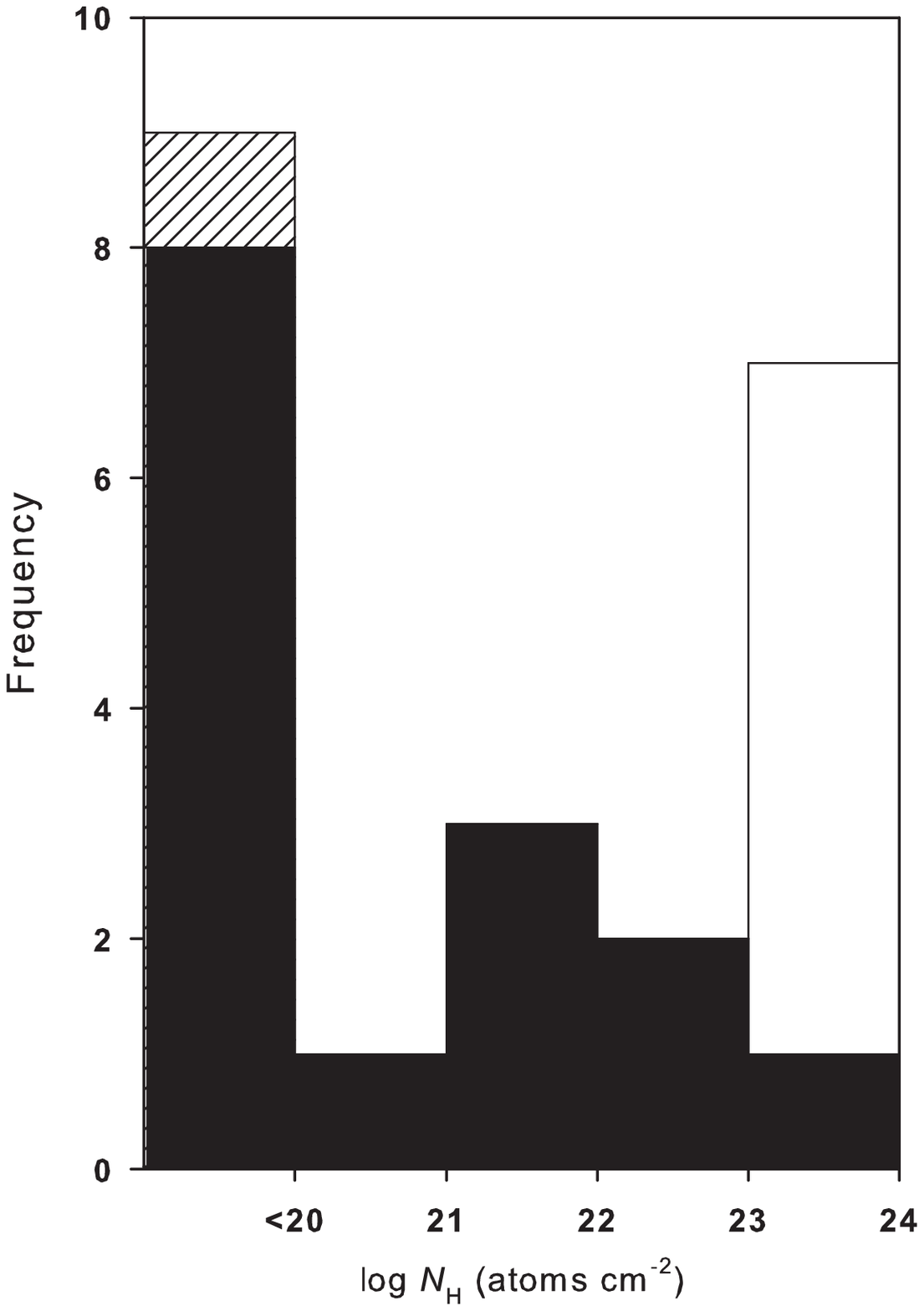}
\caption{Histogram of the intrinsic absorption associated with the
dominant component of X-ray emission in each of the sources. Black
corresponds to the FRI-type sources; white corresponds to the FRII-type
sources. The broad-line radio galaxy 3C 390.3 is (hatched box)
distinguished from the other FRII-type sources.}
\label{ch7_nhhisto}
\end{center}
\end{figure}

\begin{figure}
\begin{center}
\epsscale{1.6}
\plottwo{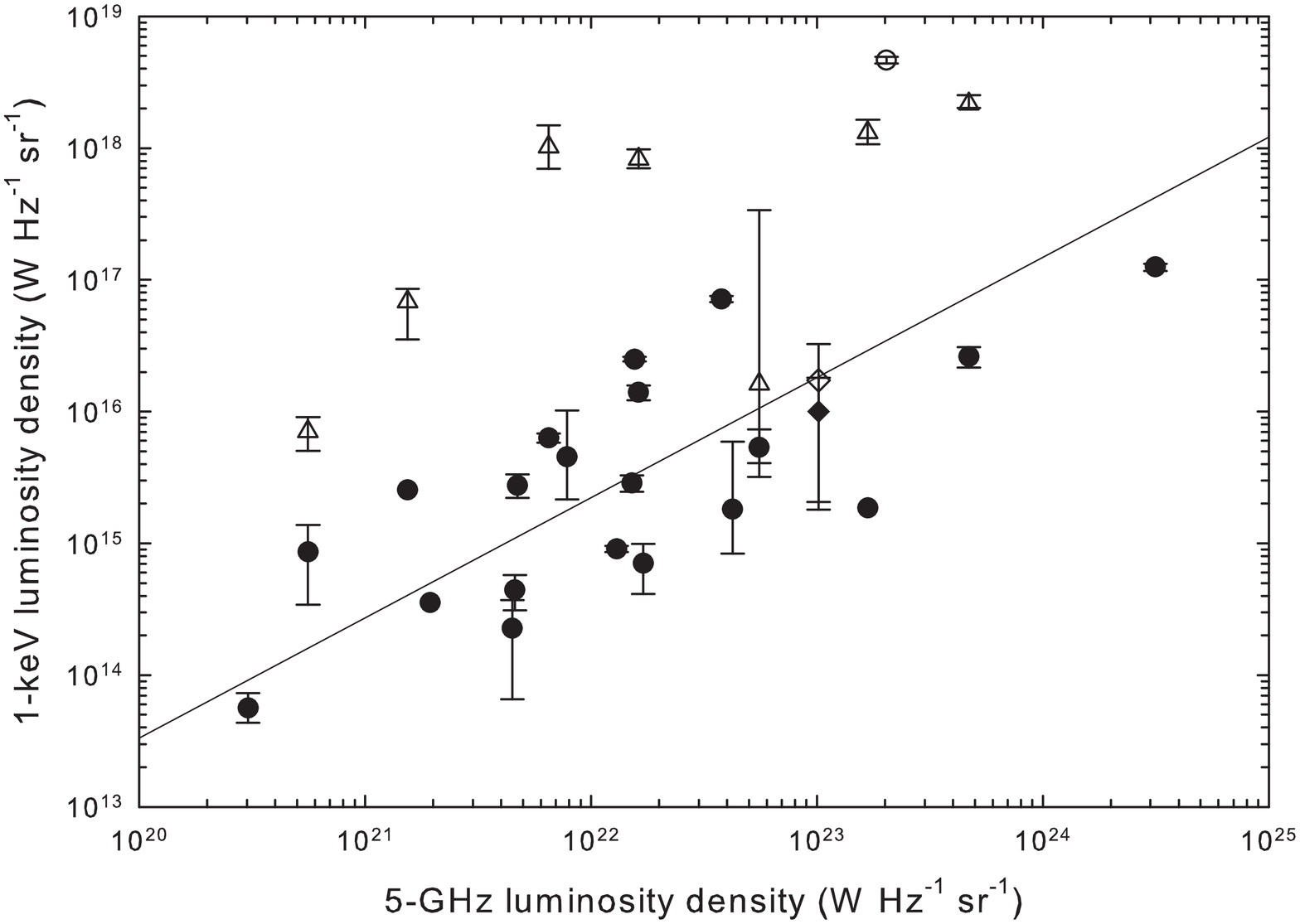}{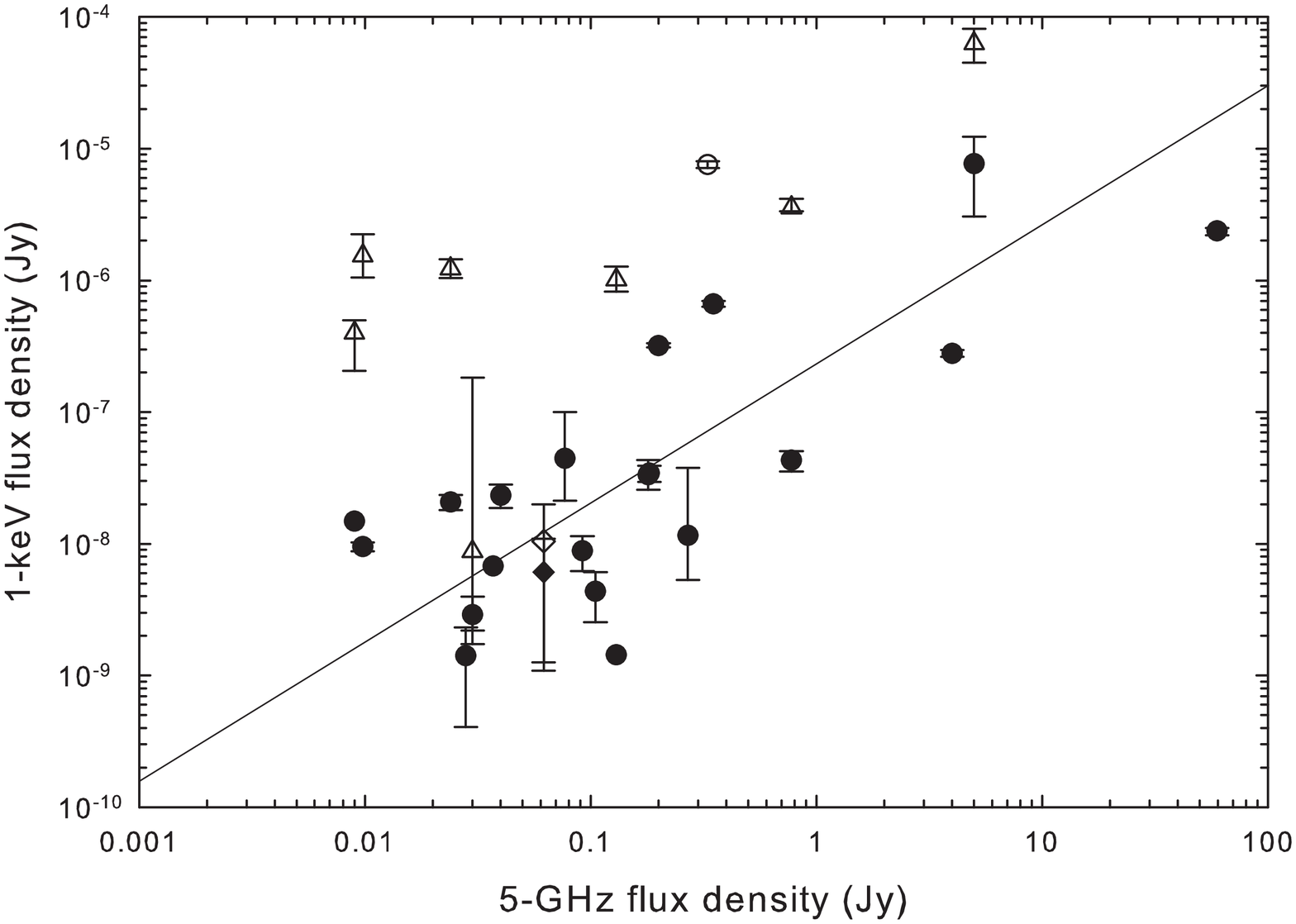}
\caption{(a) (Upper panel) 1-keV unabsorbed X-ray luminosity density against 5-GHz radio
core luminosity density for every component of X-ray emission. (b)
(Lower panel) 1-keV unabsorbed X-ray flux density against 5-GHz radio core flux
density for every component of X-ray emission. Filled
circles correspond to those components with intrinsic absorption less
than $5\times10^{22}$ atoms cm$^{-2}$; hollow triangles represent those
components with intrinsic absorption greater than $5\times10^{22}$
atoms cm$^{-2}$. The hollow circle is the broad-line radio galaxy
3C~390.3, and the two diamonds represent the low-excitation FRII-type
radio galaxy 3C~388 (see text for details). The line shown is the
bisector of the two lines of best fit obtained by the Buckley-James
regression of the 1-keV X-ray luminosity density (for components with
$N_{\rm H} < 5\times10^{22}$ atoms cm$^{-2}$) and 5-GHz radio core
luminosity density.}
\label{ch7_rxlumflux}
\end{center}
\end{figure}

\begin{figure}
\begin{center}
\epsscale{0.9}
\plotone{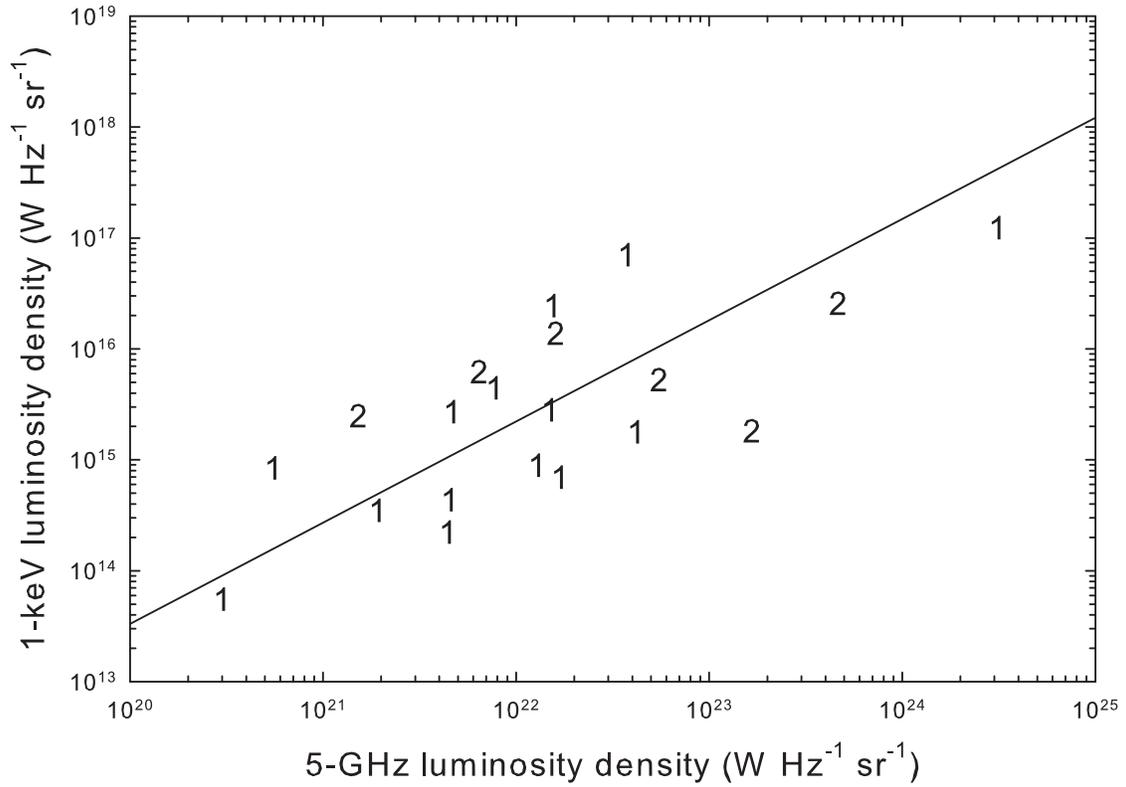}
\caption{1-keV unabsorbed X-ray luminosity density against 5-GHz radio
core luminosity density for jet-related components of X-ray emission
in FRI- and FRII-type sources. The observed X-ray and radio properties of the parsec-scale jets in FRI- and FRII-type radio galaxies are essentially indistinguishable.}
\label{ch7_fri_frii_indist_lum}
\end{center}
\end{figure}

\begin{figure}
\begin{center}
\epsscale{0.9}
\plotone{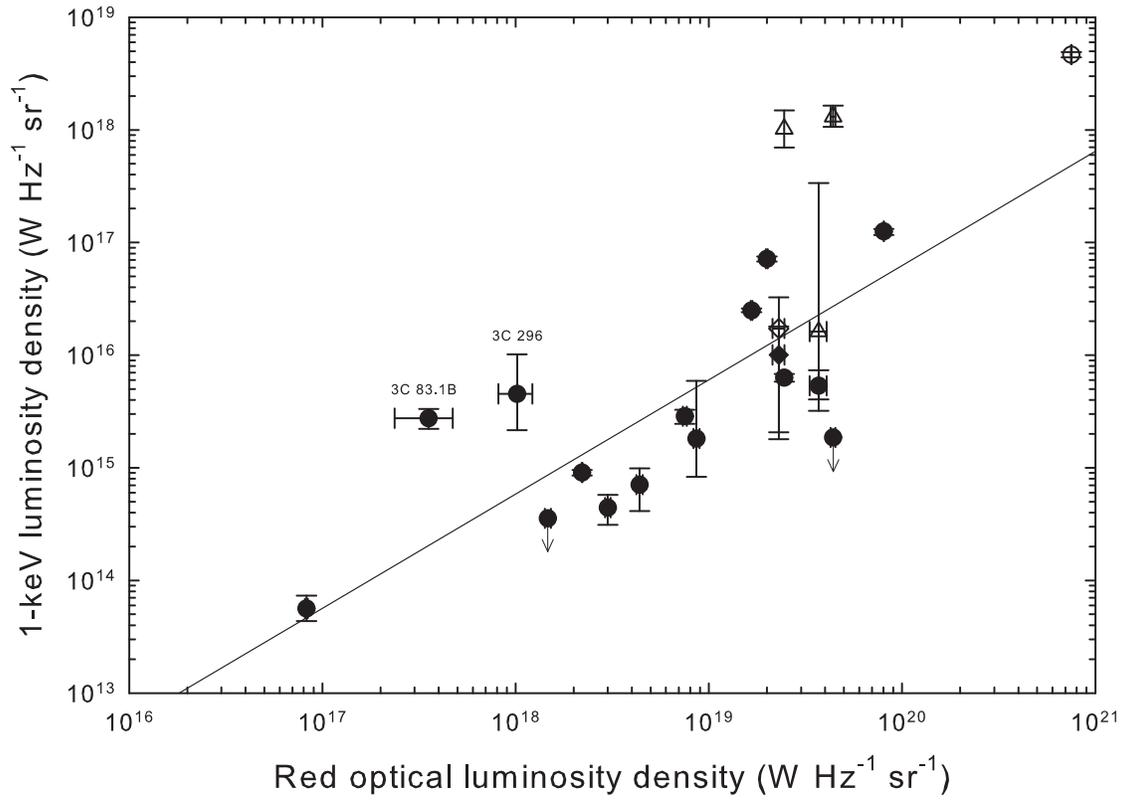}
\caption{Red optical core luminosity density (mostly using the {\it
HST} F702W filter) against 1-keV unabsorbed X-ray
luminosity density for the components studied in this paper (where the data exist). Symbols are as in Figure~\ref{ch7_rxlumflux}.}
\label{ch7_oxlum}
\end{center}
\end{figure}

\begin{figure}
\begin{center}
\epsscale{0.9}
\plotone{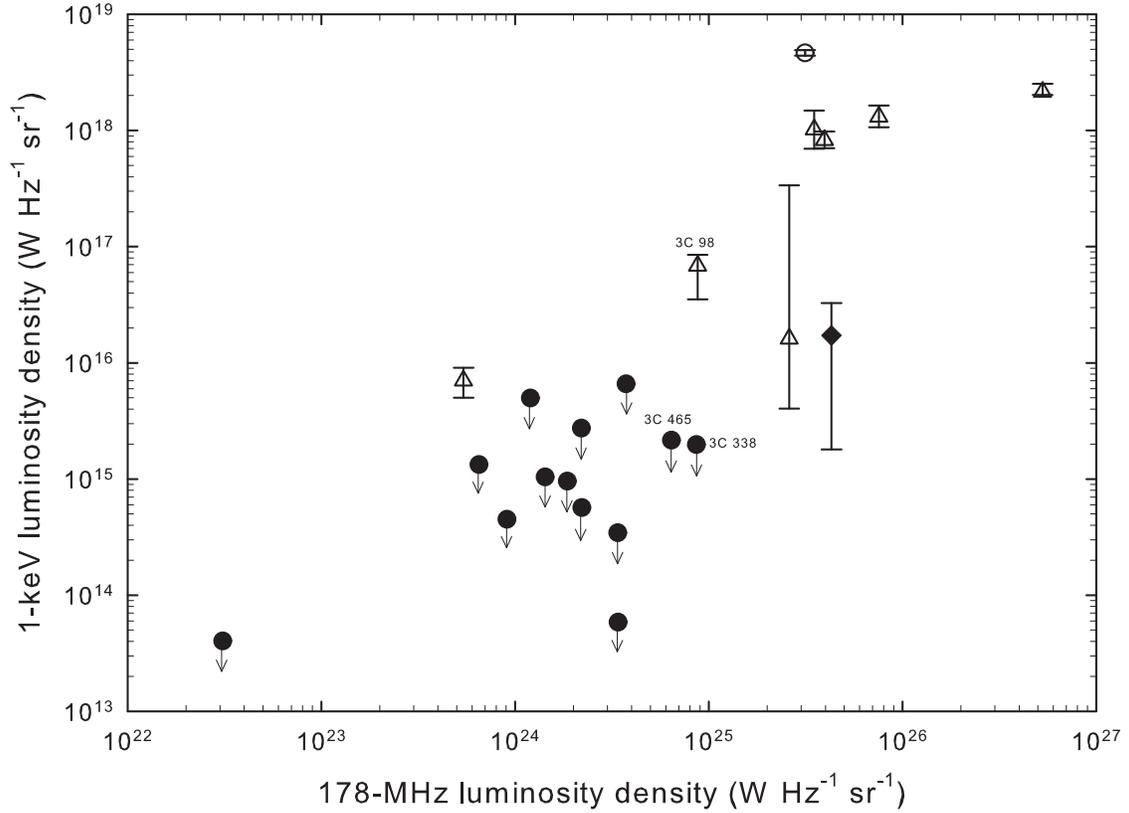}
\caption{1-keV unabsorbed luminosity density of the accretion-related
components of the sources studied in this paper against 178-MHz
luminosity density. Hollow triangles correspond to the accretion-related
emission detected in the FRII-type sources and Cen~A (the
low-excitation radio galaxy 3C~388 is depicted with a filled diamond and
the broad-line radio galaxy 3C~390.3 with a hollow circle). Filled circles correspond to the upper limits on the
luminosity densities of `hidden', accretion-related emission in the
FRI-type sources, under the assumption that the accretion flow is
obscured by a column of $10^{23}$ atoms cm$^{-2}$. Also highlighted
are 3C~98, 3C~338, and 3C~465, the three sources which populate the
FRI/FRII break in 178-MHz luminosity density.}
\label{ch7_178_corr}
\end{center}
\end{figure}

\clearpage
\begin{table}
\begin{center}
\caption{Overview of the main properties of the observed sources, ordered by 3C number.}
\label{sources_overview}
\vskip 10pt
\begin{scriptsize}
\begin{tabular}{lccccccc}
\hline

           &            & Galactic absorption &              &            &                & Optical & 178-MHz luminosity             \\
    Source &       $z$  & (atoms cm$^{-2}$)   &   FR class & RA (J2000) & DEC (J2000) &       Type & (W Hz$^{-1}$ sr$^{-1}$) \\
\hline

      3C 31              &     0.0167 & $5.41\times10^{20}$ & 1 & 01 07 24.96 & +32 24 45.21 &       LERG &   9.08$\times10^{23}$ \\
      3C 33              &     0.0595 & $4.06\times10^{20}$ & 2 & 01 08 52.86 & +13 20 13.80 &       NLRG &   3.95$\times10^{25}$ \\
     3C 66B              &     0.0215 & $8.91\times10^{20}$ & 1 & 02 23 11.41 & +42 59 31.38 &       LERG &   2.21$\times10^{24}$ \\
   3C 83.1B (NGC 1265)   &     0.0255 & $1.45\times10^{21}$ & 1 & 03 18 15.86 & +41 51 27.80 &       LERG &   3.39$\times10^{24}$ \\
      3C 84 (NGC 1275)   &     0.0177 & $1.45\times10^{21}$ & 1 & 03 19 48.16 & +41 30 42.11 &       NLRG &   3.74$\times10^{24}$ \\
      3C 98              &     0.0306 & $1.29\times10^{21}$ & 2 & 03 58 54.43 & +10 26 03.00 &       NLRG &   8.75$\times10^{24}$ \\
     3C 264              &     0.0208 & $2.45\times10^{20}$ & 1 & 11 45 05.01 & +19 36 22.74 &       LERG &   2.20$\times10^{24}$ \\
   3C 272.1 (M84)        &     0.0029 & $2.78\times10^{20}$ & 1 & 12 25 03.74 & +12 53 13.14 &       LERG &   3.1$\times10^{22}$ \\
     3C 274 (M87)        &     0.0041 & $2.54\times10^{20}$ & 1 & 12 30 49.42 & +12 23 28.04 &       NLRG &   3.4$\times10^{24}$ \\
     3C 296              &     0.0237 & $1.86\times10^{20}$ & 1 & 14 16 52.94 & +10 48 26.50 &       LERG &   1.43$\times10^{24}$ \\
     3C 321              &     0.0961 & $4.14\times10^{20}$ & 2 & 15 31 43.45 & +24 04 19.10 &       NLRG &   2.6$\times10^{25}$ \\
   NGC 6109              &     0.0296 & $1.47\times10^{20}$ & 1 & 16 17 40.54 & +35 00 15.10 &       LERG &   1.86$\times10^{24}$ \\
     3C 338              &     0.0303 & $8.90\times10^{19}$ & 1 & 16 28 38.48 & +39 33 05.60 &       NLRG &   8.63$\times10^{24}$ \\
   NGC 6251              &     0.0244 & $5.82\times10^{20}$ & 1 & 16 32 31.97 & +82 32 16.40 &       LERG &   1.2$\times10^{24}$ \\
     3C 388              &     0.0908 & $6.32\times10^{20}$ & 2 & 18 44 02.40 & +45 33 29.70 &       LERG &   4.29$\times10^{25}$ \\
   3C 390.3              &     0.0569 & $4.16\times10^{20}$ & 2 & 18 42 08.99 & +79 46 17.13 &       BLRG &   3.14$\times10^{25}$ \\
     3C 403              &     0.0590 & $1.54\times10^{21}$ & 2 & 19 52 15.80 & +02 30 24.47 &       NLRG &   3.5$\times10^{25}$ \\
     3C 405              &     0.0565 & $3.06\times10^{21}$ & 2 & 19 59 28.36 & +40 44 02.10 &       NLRG &   4.90$\times10^{27}$ \\
     3C 449              &     0.0171 & $1.19\times10^{21}$ & 1 & 22 31 20.90 & +39 21 48.00 &       LERG &   6.51$\times10^{23}$ \\
     3C 452              &     0.0811 & $1.16\times10^{21}$ & 2 & 22 45 48.77 & +39 41 15.70 &       NLRG &   7.54$\times10^{25}$ \\
     3C 465              &     0.0293 & $5.06\times10^{20}$ & 1 & 23 38 29.52 & +27 01 55.90 &       LERG &   6.41$\times10^{24}$ \\
     Cen~A               &     0.0008 & $7.69\times10^{20}$ & 1 & 13 25 27.62 & -43 01 08.81 &       NLRG &   5.4$\times10^{23}$ \\

\hline
\end{tabular}
\end{scriptsize}
\end{center}
\end{table}

\clearpage
\begin{deluxetable}{lllllllll}
\rotate
\tablewidth{23cm}
\tablecaption{Observation log}
\tabletypesize{\scriptsize}
\tablehead{&&&&& Nominal & Screened & Counts  & \\ &&& Instrument / & Observation & Exposure & Exposure & per frame & \\ Source & Telescope & Obs ID & Source CCD & Mode & (ks) & (ks) & (s$^{-1}$) & Notes \\ (1) & (2) & (3) & (4) & (5) & (6) & (7) & (8) & (9)}
\startdata
     3C 31 &        \Ch &       2147 &    ACIS-S3 &      FAINT &         50 &       43.3 &       0.06 &            \\

     3C 33 &       \XMM & 0203280301 &       EPIC & MEDIUM+MEDIUM+MEDIUM &     7 (pn) &   6.3 (pn) &  1.20 (pn) &            \\
 
    3C 66B &        \Ch &        828 &    ACIS-S3 &      FAINT &         50 &       29.6 &       0.06 &            \\
 
  3C 83.1B &        \Ch &       3237 &    ACIS-S3 &     VFAINT &         95 &       84.5 &       0.02 &            \\
 
     3C 84 &       \XMM & 00085110101 &       EPIC & THIN+THIN+THIN &    51 (pn) &  24.7 (pn) & 14.25 (pn) &            \\
 
     3C 98 &       \XMM & 0064600301 &       EPIC & THICK+THICK+MEDIUM &    10 (pn) &   3.2 (pn) &   0.3 (pn) &            \\
 
    3C 264 &        \Ch &       4916 &    ACIS-S3 &      FAINT &         40 &       34.4 &       0.14 &            \\
 
  3C 272.1 &        \Ch &        803 &    ACIS-S3 &     VFAINT &         30 &       28.1 &       0.09 &            \\
 
    3C 274 &        \Ch &       1808 &    ACIS-S3 &      FAINT &         14 &       12.8 &       0.21 &            \\
 
    3C 296 &        \Ch &       3968 &    ACIS-S3 &     VFAINT &         50 &       48.6 &       0.07 &            \\
 
    3C 321 &        \Ch &       3138 &    ACIS-S3 &      FAINT &         50 &       46.6 &       0.04 &            \\
 
  NGC 6109 &        \Ch &       3985 &    ACIS-S3 &     VFAINT &         20 &       19.0 &       0.02 &            \\
 
    3C 338 &        \Ch &        497 &    ACIS-S3 &      FAINT &         20 &       18.3 &       0.05 &            \\
 
  NGC 6251 &        \Ch &       4130 &    ACIS-I3 &     VFAINT &         50 &       43.0 &       0.36 & See \cite{evans05} \\
 
    3C 388 &        \Ch &       5295 &    ACIS-I3 &     VFAINT &         33 &       30.7 &       0.02 &            \\
 
  3C 390.3 &        \Ch &        830 &    ACIS-S3 &      FAINT &         35 &       23.9 &         -- & Readout streak \\
 
    3C 403 &        \Ch &       2968 &    ACIS-S3 &      FAINT &         50 &       45.9 &       0.14 & See \cite{kra05}           \\
 
    3C 405 &        \Ch &       1707 &    ACIS-S3 &     VFAINT &         10 &        9.2 &       0.06 &            \\
 
    3C 449 &        \Ch &       4057 &    ACIS-S3 &     VFAINT &         30 &       24.7 &      0.035 &            \\
 
    3C 449 &       \XMM & 0002970101 &       EPIC & MEDIUM+MEDIUM+MEDIUM &  18.5 (pn) &  16.7 (pn) &  0.35 (pn) &            \\
 
    3C 452 &        \Ch &       2195 &    ACIS-S3 &      FAINT &         80 &       79.5 &       0.09 &            \\
 
    3C 465 &        \Ch &       4816 &    ACIS-S3 &     VFAINT &         50 &       49.3 &       0.02 &            \\
 
    3C 465 &       \XMM & 0002960101 &       EPIC & MEDIUM+MEDIUM+MEDIUM &   9.7 (pn) &   7.8 (pn) &  0.85 (pn) &            \\
 
     Cen A &        \Ch & 1600/1601  &     HETGS  &      FAINT &        100 &       98.3 &         -- & See \cite{evans04}           \\

     Cen A &       \XMM & 0093650201 &       EPIC & MEDIUM+MEDIUM+MEDIUM &  19.3 (pn) &  16.8 (pn) & 15 (pn) & See \cite{evans04} \\

     Cen A &       \XMM & 0093650301 &       EPIC & MEDIUM+MEDIUM+MEDIUM &   9.3 (pn) &   7.9 (pn) & 15 (pn) & See \cite{evans04} \\ 

\enddata
\label{obslog}
\tablecomments{Col. (1): Name of source. Col. (2): Telescope used to perform observation. Col. (3): Observation ID. Col. (4): Instrument / CCD source located on. Col. (5): Observation mode. Indicates data mode for \Ch and optical blocking filter for \XMM MOS1, MOS2, and pn cameras. Col. (6): Nominal exposure time. Col. (7): Screened exposure time. Col. (8): Point source counts per frame time. Used as a pileup diagnostic. Col. (9): Additional remarks.}
\end{deluxetable}

\clearpage
\begin{deluxetable}{lllllllllll}
\rotate
\tablewidth{26cm}
\tablecaption{Main spectral parameters of the $z<0.1$ radio-galaxy nuclei}
\tabletypesize{\tiny}
\tablehead{           &             &                                &            &            &                 &        &    & $L_{\rm (2-10 keV)}$                         &        &            \\
           &             &                                &            &            &                 &        &    & (Power Law)                                  &        &            \\
    Source & Telescope   &   Description of best spectrum &  $N_{\rm H}$ (atoms cm$^{-2}$) &   $\Gamma$ &    E (keV) &  $\sigma$ (keV) & $kT$ (keV) & (ergs s$^{-1}$) &  $\chi^{2}$/dof &   Comments \\
    (1)    &   (2)       &             (3)                &               (4)              &     (5)    &     (6)    &      (7)        &      (8)   &             (9) &          (10)   &      (11)}
\startdata
     3C 31 & C           & PL+TH &       -- & $1.48^{+0.28}_{-0.32}$ &         -- &         -- & $0.68^{+0.08}_{-0.07}$ & 4.7$\times10^{40}$ &    9.6/25 &         -- \\
 
     3C 33 & X           & $N_{\rm H}$(PL+Gauss)+PL & $(3.9^{+0.7}_{-0.6})\times10^{23}$ & $\Gamma_1 = 1.7$ (f); & 6.42 $\pm$ 0.09 &    0.1 (f) &         -- & 6.3$\times10^{43}$ &   49.0/39 &
    -- \\
           &             &                &            --                         & $\Gamma_2 = 1.7$ (f)  &                 &            &                    & 1.7$\times10^{42}$ &           &
        \\
 
    3C 66B & C         &      PL+TH &         -- & $2.03\pm0.18$ &         -- &         -- &$0.41^{+0.23}_{-0.10}$ & 1.3$\times10^{41}$ &   29.1/34 &            \\
 
  3C 83.1B & C         & $N_{\rm H}$(PL)+TH & $(3.2^{+0.8}_{-0.7})\times10^{22}$ & $2.00^{+0.27}_{-0.20}$ &         -- &         -- & $0.58^{+0.15}_{-0.14}$ & 1.3$\times10^{41}$ &    9.5/12 &
       \\
 
     3C 84 & X           & PL+Gauss+TH+TH &         -- & $1.81 \pm 0.03$ & $6.39^{+0.08}_{-0.09}$ & $ 0.01$ (f)      & $kT_1 = 0.77 \pm 0.04$;  & 8.2$\times10^{42}$ & 1156.0/1151 &  \\
           &             &                 &            &                 &                        &                 & $kT_2 = 2.74 \pm 0.10       $ &                &              &
                                    \\
     3C 98 & X           & $N_{\rm H}$ (PL+Gauss)+TH & $(1.2^{+0.3}_{-0.2})\times10^{23}$ & $1.68^{+0.23}_{-0.37}$ & 6.37 $\pm$ 0.10 &    0.1 (f) & $0.98 \pm 0.12$ & 5.4$\times10^{42}$ &   32.8/40 &            \\
 
    3C 264 & C         &         PL &         -- & $2.34^{+0.07}_{-0.08}$ &         -- &         -- &         -- & 7.5$\times10^{41}$ & 123.0/147 &            \\
 
  3C 272.1 & C         & $N_{\rm H}$(PL) & $(1.9^{+0.8}_{-0.7})\times10^{21}$ & $2.14^{+0.34}_{-0.30}$ &         -- &         -- &         -- & 2.2$\times10^{39}$ &   11.3/24 &            \\
 
    3C 274 & C         & PL+TH &         -- & $2.09 \pm 0.06$ &         -- &         -- & $0.75^{+0.17}_{-0.14}$ & 3.9$\times10^{40}$ &   84.2/98 &            \\
 
    3C 296 & C         & $N_{\rm H}$ (PL)+TH & $(1.5\pm0.9)\times10^{22}$ & $1.77^{+0.60}_{-0.52}$ &         -- &         -- & $0.75^{+0.27}_{-0.46}$ & 3.1$\times10^{41}$ &    9.3/21 &
\\
 
    3C 321 & C         & $N_{\rm H}$(PL+Gauss)+PL+TH & $(1.5^{+9.6}_{-0.9})\times10^{23}$ & $\Gamma_1 = 1.7$ (f); &   6.40 (f) &    0.5 (f) & $0.49 \pm 0.15$ & $(1.2^{+24.3}_{-0.9})\times10^{42}$
&    9.8/8 &  \\
           &             &                        &   --                                   & $\Gamma_2 = 2$ (f)    &            &            &                & $(2.6^{+1.0}_{-1.1})\times10^{41}$
 &            & \\
 
  NGC 6109 & C         & PL+TH &         -- & $1.47 \pm 0.47$ &         -- &         -- & $0.63^{+0.14}_{-0.17}$ & 2.3$\times10^{40}$ &     1.5/6 &            \\
 
    3C 338 & C         &         PL &         -- & $2.37 \pm 0.81$ &         -- &         -- &         -- & 2.0$\times10^{40}$ &     4.5/3 &            \\
 
  NGC 6251 & C         & $N_{\rm H}$(PL)+TH & $4.5\times10^{20}$ (f) & $1.67 \pm 0.06$ &         -- &         -- & $0.20 \pm 0.08$ & 5.9$\times10^{42}$ &    107/134 &            \\
 
    3C 388 & C         &   PL+TH    &         -- & 2 (f)           &         -- &         -- & 1.03 $\pm$ 0.29 & $4.9\times10^{41}$ & 6.4/8 &  \\
 
  3C 390.3 & C         &         PL &         -- & $1.78 \pm 0.09$ &         -- &         -- &         -- & 3.2$\times10^{44}$ &   24.9/54 & Tentative Fe K$\alpha$ line \\
                                                                                                                                                                                                    
    3C 449 & C         &         PL &         -- & $1.67^{+0.45}_{-0.49}$ &         -- &         -- &         -- & $<$2.9$\times10^{40}$ &     8.8/4 & Upper limits \\
                                                                                                                                                                                                    
     3C 403 & C           & $N_{\rm H}$(PL+Gauss)+PL & $(4.5^{+0.7}_{-0.6})\times10^{23}$ & $\Gamma_1 = 1.76\pm0.23$; & 6.32 $\pm$ 0.02 &    0.1 (f) &         -- & 7.1$\times10^{43}$ &   31.1/50 &         -- \\
           &             &                &               --                      & $\Gamma_2 = 2$ (f)  &                 &            &                    & 3.1$\times10^{41}$ &           &
      \\
                                                                                                                                                                                                    
     3C 405 & C           & $N_{\rm H}$(PL+Gauss)+PL & $(1.7^{+0.4}_{-0.3})\times10^{23}$ & $\Gamma_1 = 1.60^{+0.53}_{-0.54}$; & 6.41 $\pm$ 0.04 &    0.1 (f) &         -- & 1.9$\times10^{44}$ &
69.4/84 &         -- \\
           &             &                &                    --                 & $\Gamma_2 = 2$ (f)  &                 &            &                    & 1.3$\times10^{42}$ &           &
      \\
                                                                                                                                                                                                    
    3C 452 & C         & $N_{\rm H}$(PL+Gauss)+PEXRAV+TH & $(5.7^{+0.9}_{-0.8})\times10^{23}$ &    1.7 (f) &    6.4 (f) &    0.1 (f) & $0.63 \pm 0.29$ & 1.0$\times10^{44}$ &   65.9/69 & Follows \cite{isobe02} \\
                                                                                                                                                                                                    
    3C 465 & C         & $N_{\rm H}$(PL)+TH & $(4.5^{+6.0}_{-3.9})\times10^{21}$ & $1.86^{+0.72}_{-0.52}$ &         -- &         -- & $0.70^{+0.07}_{-0.06}$ & 1.1$\times10^{41}$ &   22.2/23 &
       \\
                                                                                                                                                                                                    
     Cen A & C/X         & $N_{\rm H}$(PL+Gauss)+$N_{\rm H}$(PL) & $N_{\rm H,1} = (1.2\pm0.2)\times10^{23}$ & $\Gamma_1 = 1.72\pm0.21$; & 6.40 $\pm$ 0.01 & $0.02 \pm 0.01$    &         -- & $\sim5\times10^{41}$ &   -- &     See \cite{evans04} \\
           &             &                &       $N_{\rm H,2} = (3.8\pm2.0)\times10^{22}$      & $\Gamma_2 = 2$ (f)  &                 &            &                    & 2.8$\times10^{37}$ &           &
      \\

\enddata
\label{3crr_spectral results}
\tablecomments{Col. (1): Name of source. Col. (2): Telescope used (C={\it
Chandra}, X={\it XMM-Newton}). Col. (3): Description of best spectrum
($N_{\rm H}$=Intrinsic absorption, PL=Power Law, Gauss=Redshifted
Gaussian Line, TH=Thermal, PEXRAV=Reflection from neutral
material. Col. (4): Intrinsic neutral hydrogen column
density. Galactic absorption has also been applied (see
Table~\ref{sources_overview} for values). Col. (5): Power-law photon index. Col. (6): Gaussian centroid energy. Col. (7): Gaussian linewidth. Col. (8): Thermal temperature. Col. (9): 2--10 keV unabsorbed luminosity of primary power law. Col. (10): Value of $\chi^{2}$ and degrees of freedom. Col. (11): Comments. (f): Indicates parameter was frozen.}
\end{deluxetable}

\clearpage
\begin{deluxetable}{llllllllll}
\rotate
\tablewidth{22cm}
\tablecaption{X-ray, radio, and optical flux and luminosity densities}
\tabletypesize{\tiny}
\tablehead{&&& 5-GHz &&& 5-GHz &&\\&& 1-keV flux & VLA flux & HST flux & 1-keV luminosity & VLA luminosity &   HST luminosity & \\ Source & $N_{\rm H}$ (atoms cm$^{-2}$) & density (nJy) & density (Jy) & density ($\mu$Jy) & density (W Hz$^{-1}$ s$^{-1}$r) & density (W Hz$^{-1}$ sr$^{-1}$) & density (W Hz$^{-1}$ sr$^{-1}$) & HST Reference  \\ (1) & (2) & (3) & (4) & (5) & (6) & (7) & (8) & (9)}
\startdata
     3C 31 &         -- & $8.9 \pm 2.6$ &       0.09 & $60 \pm 2$ & $(4.4 \pm 1.3)\times10^{14}$ & $4.60\times10^{21}$ & $(3.0 \pm 0.1)\times10^{18}$ & HW00 \\ 

     3C 33 & $(3.9^{+0.7}_{-0.6})\times10^{23}$ & $1200\pm200$ & 0.02\tablenotemark{\ast} & -- & $(8.3^{+1.5}_{-1.3})\times10^{17}$ & $1.62\times10^{22}$\tablenotemark{\ast} & -- & -- \\

           &         -- & $21 \pm 3$ &            &            & $(1.4 \pm 0.2)\times10^{16}$ &            &            &            &            \\

    3C 66B &         -- & $34 \pm 5$ &       0.18 & $90 \pm 2$ & $(2.9 \pm 0.4)\times10^{15}$ & $1.52\times10^{22}$ & $(7.5 \pm 0.2)\times10^{18}$ &     HW00 \\ 

  3C 83.1B & $(3.2^{+0.8}_{-0.7})\times10^{22}$ & $23\pm5$ &       0.04 & $3 \pm 1$ & $(2.8^{+0.6}_{-0.5})\times10^{15}$ & $4.72\times10^{21}$ & $(3.5 \pm 1.2)\times10^{17}$ &     HW00 \\ 

     3C 84 &         -- & $2400^{+100}_{-200}$ &      59.60 & $1500 \pm 10$ & $(1.3 \pm 0.1)\times10^{17}$ & $3.15\times10^{24}$ & $(8.0 \pm 0.01)\times10^{19}$ &     HW00 \\ 
     3C 98 & $(1.2^{+0.3}_{-0.2})\times10^{23}$ & $400^{+100}_{-190}$ & 0.01\tablenotemark{\ast} & -- & $(6.9^{+1.7}_{-3.3})\times10^{16}$ & $1.54\times10^{21}$\tablenotemark{\ast} & -- & -- \\
           &         -- & $<15$ (fs) &            &            & $<2.5\times10^{15}$ (fs) &            &            &            &            \\ 

    3C 264 &         -- & $320\pm10$ &       0.20 & $210 \pm 2$ & $(2.5^{+0.1}_{-0.8})\times10^{16}$ & $1.56\times10^{22}$ & $(1.7 \pm 0.02)\times10^{19}$ &     HW00 \\ 

  3C 272.1 & $(1.9^{+0.8}_{-0.7})\times10^{21}$ & $33^{+10}_{-8}$ &       0.18 & $49 \pm 1$ & $(5.6^{+1.7}_{-1.3})\times10^{13}$ & $3.04\times10^{20}$ & $(8.3 \pm 0.2)\times10^{16}$ &     HW00 \\ 

    3C 274 &         -- & $280 \pm 20$ &       4    & $680 \pm 2$ & $(9.1 \pm 0.5)\times10^{14}$ & $1.30\times10^{22}$ & $(2.2 \pm 0.01)\times10^{18}$ &     HW00 \\ 

    3C 296 & $(1.5\pm0.9)\times10^{22}$ & $45^{+56}_{-23}$ &       0.08 & $10 \pm 2$ & $(4.5^{+5.6}_{-2.4})\times10^{15}$ & $7.83\times10^{21}$ & $(1.0 \pm 0.2)\times10^{18}$ &     HW00 \\ 
    3C 321 & $(1.5^{+9.6}_{-0.9})\times10^{23}$ & $8.8^{+174}_{-6.6}$ & 0.03\tablenotemark{\ast} & $20 \pm 2$\tablenotemark{\ast} & $(1.6^{+32}_{-1.2})\times10^{16}$ & $5.55\times10^{22}$\tablenotemark{\ast} & $(3.7 \pm 0.4)\times10^{19}$\tablenotemark{\ast} & SH04 \\

           &         -- & $2.9^{+1.1}_{-1.2}$ &            &            & $(5.4^{+2.0}_{-2.2})\times10^{15}$ &            &            &            &            \\ 
 
  NGC 6109 &         -- & $1.4^{+0.9}_{-1.0}$ &       0.03 &         -- & $(2.3^{+1.5}_{-1.6})\times10^{14}$ & $4.5\times10^{21}$ &         -- &         -- \\ 

    3C 338 &         -- & $4.4^{+1.7}_{-1.8}$ &       0.11 & $27 \pm 1$ & $(7.1^{+2.8}_{-2.9})\times10^{14}$ & $1.70\times10^{22}$ & $(4.4 \pm 0.2)\times10^{18}$ &     HW00 \\ 

  NGC 6251 & $4.5\times10^{20}$ (f) & $660 \pm 30$ &       0.35 & $190 \pm 2$ & $(7.2 \pm 0.04)\times10^{16}$ & $3.78\times10^{22}$ & $(2.0 \pm 0.02)\times10^{19}$ & \cite{evans05} \\ 

    3C 388 &         -- & $6.1\pm4.9$ &               0.06   & $14 \pm 1$  & $(1.0\pm0.8)\times10^{16}$ & $1.0\times10^{23}$ & $(2.3 \pm 0.2)\times10^{19}$ & SH04 \\ 

  3C 390.3 &         -- & $7600 \pm 400$ &       0.33 & $1200 \pm 10$ & $(4.7 \pm 0.3)\times10^{18}$ & $2.30\times10^{23}$ & $(7.5 \pm 0.02)\times10^{20}$ &     HW00 \\ 

    3C 403 & $(4.5^{+0.7}_{-0.6})\times10^{23}$ & $1500^{+700}_{-500}$ & $9.8\times10^{-3}$\tablenotemark{\ast}    & 37\tablenotemark{\ast} & $(1.0^{+0.5}_{-0.3})\times10^{18}$ & $6.50\times10^{21}$\tablenotemark{\ast} & $2.5\times10^{19}$\tablenotemark{\ast} & KS04 \\

           & --  & $9.5 \pm 0.8$ &            &            & $(6.3 \pm 0.5)\times10^{15}$ &            &            &            &            \\ 

    3C 405 & $(1.7^{+0.4}_{-0.3})\times10^{23}$ & $3600^{+600}_{-200}$ & 0.78\tablenotemark{\ast}    & -- & $(2.2^{+0.4}_{-0.1})\times10^{18}$ & $4.7\times10^{23}$\tablenotemark{\ast} & -- & -- \\

           & --  & $43 \pm 8$ &            &            & $(2.6 \pm
0.5)\times10^{16}$ &            &            &            &
\\ 

    3C 449 &         -- & $<6.8$ &       0.04 & $28 \pm 1$ &  $<3.6\times10^{14}$ & $1.94\times10^{21}$ & $(1.5 \pm 0.05)\times10^{18}$ &     HW00 \\ 

    3C 452 & $(5.7^{+0.9}_{-0.8})\times10^{23}$ & $1000^{+300}_{-200}$ & 0.13\tablenotemark{\ast} & $34 \pm 1$\tablenotemark{\ast} & $(1.31\pm0.3)\times10^{18}$ & $1.68\times10^{23}$\tablenotemark{\ast} & $(4.4 \pm 0.1)\times10^{19}$\tablenotemark{\ast} & SH04 \\

           &         -- & $<1.44$ (fs) &            &         -- & $<1.86\times10^{15}$ (fs) &            &            &            &            \\ 

    3C 465 & $(4.5^{+6.0}_{-3.9})\times10^{21}$ & $12^{+26}_{-6}$ &       0.27 & $55 \pm 2$ & $(1.8^{+4.1}_{-1.0})\times10^{15}$ & $4.23\times10^{22}$ & $(8.6 \pm 0.3)\times10^{18}$ &     HW00 \\

     Cen A & $(1.2 \pm 0.1)\times10^{23}$ & $63000 \pm 18000$ & 5\tablenotemark{\ast}    & -- & $(7.1 \pm 2.0)\times10^{15}$ & $5.60\times10^{20}$\tablenotemark{\ast} & -- & -- \\

           & $(3.6^{+2.2}_{-2.3})\times10^{22}$ & $7700 \pm 4600$ &            &            & $(8.6 \pm 5.2)\times10^{14}$ &            &            &            &            \\ 

\enddata
\label{3crr_xro_flux_lum_tab}
\tablenotetext{\ast}{Radio/optical flux/luminosity density quoted is total and is not implicitly associated with this component of X-ray emission}
\tablecomments{Col. (1): Name of source. Col. (2): Intrinsic absorption of
X-ray component. Col. (3): 1-keV unabsorbed flux density of X-ray
component. Col. (4): 5-GHz VLA flux density of core. Col. (5): Red
{\it HST} flux density (dereddened for Milky Way). Mostly uses data
from the F702W filter. Col. (6): 1-keV unabsorbed
luminosity density of X-ray component. Col. (7): 5-GHz VLA luminosity
density of core. Taken from the online 3CRR catalogue (http://www.3crr.dyndns.org/) and references
therein. Col. (8): Red {\it HST} luminosity density (dereddened for Milky Way). Mostly uses data
from the F702W filter. Col. (9): Reference used for HST data (HW00=\cite{har00}, KS04=\cite{ks04}, SH04=O. Shorttle, private communication. (fs): Upper limit to soft, unabsorbed X-ray emission in heavily absorbed sources (not statistically required).}
\end{deluxetable}

\begin{table}[h]
\begin{center}
\caption{Black hole masses, and unabsorbed X-ray and Eddington
luminosities and efficiencies for components with $N_{\rm H} >
5\times10^{22}$ atoms cm$^{-2}$ plus the BLRG 3C~390.3}
\label{ch7_eddingtonefficiencies}
\vskip 10pt
\begin{tabular}{lcccc}
\hline
Source & log $M_{\rm BH}$ (M$_\odot$) & $L_{\rm Edd}$ (ergs s$^{-1}$) & $L_{\rm 0.5-10 keV}$ (ergs s$^{-1}$) & $\eta_{\rm X,Edd}$ \\
\hline
     3C 33 &       8.68 &    6.2$\times10^{46}$ &    9.7$\times10^{43}$ &    1.6$\times10^{-3}$ \\
     3C 98 &       8.23 &    2.2$\times10^{46}$ &    8.2$\times10^{42}$ &    3.7$\times10^{-4}$ \\
  3C 390.3 &       8.53 &    4.4$\times10^{46}$ &    5.0$\times10^{44}$ &    1.1$\times10^{-2}$ \\
    3C 403 &       8.41 &    3.3$\times10^{46}$ &    1.1$\times10^{44}$ &    3.3$\times10^{-3}$ \\
    3C 405 &       9.40 &    3.3$\times10^{47}$ &    2.8$\times10^{44}$ &    8.5$\times10^{-4}$ \\
    3C 452 &       8.54 &    4.5$\times10^{46}$ &    1.5$\times10^{44}$ &    3.3$\times10^{-3}$ \\
     Cen A &       8.30 &    2.6$\times10^{46}$ &    7.8$\times10^{41}$ &    3.0$\times10^{-5}$ \\
\hline
\end{tabular}
\end{center}
\end{table}

\clearpage
\begin{deluxetable}{lcccccc}
\rotate
\tablewidth{17cm}
\tablecaption{90\%-confidence upper limits to hidden accretion-related emission in likely jet-dominated FRI-type radio-galaxy nuclei, assuming obscuring columns of either $10^{23}$ atoms cm$^{-2}$ or $10^{24}$ atoms cm$^{-2}$. Shown are the black hole masses, unabsorbed 0.5--10 keV X-ray and Eddington luminosities, and efficiencies.}
\tabletypesize{\scriptsize}
\tablehead{       &                              &                               & \multicolumn{2}{c}{$N_{\rm H} = 10^{23}$ atoms cm$^{-2}$} & \multicolumn{2}{c}{$N_{\rm H} = 10^{24}$ atoms cm$^{-2}$} \\
Source & log $M_{\rm BH}$ (M$_\odot$) & $L_{\rm Edd}$ (ergs s$^{-1}$) & $L_{\rm 0.5-10 keV}$ (ergs s$^{-1}$) & $\eta_{\rm X,Edd}$ & $L_{\rm 0.5-10 keV}$ (ergs s$^{-1}$) & $\eta_{\rm X,Edd}$}
\startdata
     3C 31 & 7.89 & 1.0$\times10^{46}$ & $<$5.4$\times10^{40}$ & $<$5.4$\times10^{-6}$ & $<$2.0$\times10^{42}$ & $<$2.0$\times10^{-4}$ \\
    3C 66B & 8.84 & 9.0$\times10^{46}$ & $<$6.7$\times10^{40}$ & $<$7.4$\times10^{-7}$ & $<$4.0$\times10^{42}$ & $<$4.4$\times10^{-5}$ \\
  3C 83.1B & 9.01 & 1.3$\times10^{47}$ & $<$4.1$\times10^{40}$ & $<$3.1$\times10^{-7}$ & $<$8.8$\times10^{41}$ & $<$6.8$\times10^{-6}$\\
     3C 84 & 9.28 & 2.5$\times10^{47}$ & $<$7.8$\times10^{41}$ & $<$3.1$\times10^{-6}$ & $<$2.3$\times10^{42}$ & $<$9.2$\times10^{-6}$ \\
    3C 264 & 8.85 & 9.2$\times10^{46}$ & $<$3.3$\times10^{41}$ & $<$3.6$\times10^{-6}$ & $<$1.7$\times10^{42}$ & $<$1.8$\times10^{-5}$ \\
  3C 272.1 & 9.18 & 1.9$\times10^{47}$ & $<$4.8$\times10^{39}$ & $<$2.5$\times10^{-8}$ & $<$1.6$\times10^{41}$ & $<$8.5$\times10^{-7}$ \\
    3C 274 & 9.38 & 3.0$\times10^{47}$ & $<$7.0$\times10^{39}$ & $<$2.3$\times10^{-8}$ & $<$1.3$\times10^{41}$ & $<$4.3$\times10^{-7}$ \\
    3C 296 & 9.13 & 1.8$\times10^{47}$ & $<$1.2$\times10^{41}$ & $<$6.8$\times10^{-7}$ & $<$2.1$\times10^{42}$ & $<$1.2$\times10^{-5}$ \\
  NGC 6109 &   -- &      --            & $<$1.1$\times10^{41}$ &       --              & $<$2.1$\times10^{42}$ & --  \\
    3C 338 & 9.23 & 2.2$\times10^{47}$ & $<$2.3$\times10^{41}$ & $<$1.0$\times10^{-6}$ & $<$4.4$\times10^{42}$ & $<$2.0$\times10^{-5}$ \\
  NGC 6251 & 8.78 & 7.8$\times10^{46}$ & $<$5.9$\times10^{41}$ & $<$7.6$\times10^{-6}$ & $<$1.6$\times10^{43}$ & $<$2.0$\times10^{-4}$ \\
    3C 449 & 7.71 & 6.7$\times10^{45}$ & $<$1.6$\times10^{41}$ & $<$2.4$\times10^{-5}$ & $<$4.7$\times10^{43}$ & $<$7.0$\times10^{-3}$ \\
    3C 465 & 9.32 & 2.7$\times10^{47}$ & $<$2.6$\times10^{41}$ & $<$9.6$\times10^{-7}$ & $<$5.9$\times10^{43}$ & $<$2.2$\times10^{-4}$ \\

\enddata
\label{ch7_freezepow1_tab}
\end{deluxetable}

\end{document}